\documentclass{article}
\usepackage{arxiv}
\usepackage{amsmath,amssymb,amsfonts}
\usepackage{float}
\usepackage{placeins}
\usepackage{soul}
\usepackage{algorithm}
\usepackage{graphicx}
\usepackage{textcomp}
\usepackage{comment}
\usepackage{booktabs}
\usepackage{pifont}

\usepackage{subfiles}
\usepackage[table,xcdraw]{xcolor}
\usepackage{multirow}
\usepackage{array}
\usepackage{listings}
\usepackage{ifthen}
\usepackage{fancyhdr}
\usepackage{soul}
	\makeatletter \def\@fancyvbox#1#2{\vbox{#2}} \makeatother
\usepackage[font=small,format=plain,labelfont= sc]{caption}
\usepackage[utf8]{inputenc}
\usepackage[T1]{fontenc}
\usepackage{color}
\usepackage{hyperref}
\usepackage{booktabs,multirow}
\usepackage{microtype}
\usepackage{tikz,bibunits,graphicx,circuitikz}
\usepackage{paralist}
\usepackage{pgfplots, pgfplotstable}
\usepgfplotslibrary{statistics}
\usetikzlibrary{patterns,intersections}
\pgfplotsset{compat=newest}
\usepackage{makecell}
\usepackage{tablefootnote}
\usepackage{float,placeins}
\usepackage{cellspace}
\usepackage{longtable}
\usepackage{xspace}
\usepackage{colortbl}
\usepackage{makecell} 

\newcolumntype{M}[1]{>{\centering\arraybackslash}m{#1}} 

\def\BibTeX{{\rm B\kern-.05em{\sc i\kern-.025em b}\kern-.08em
T\kern-.1667em\lower.7ex\hbox{E}\kern-.125emX}}

\hypersetup{
    colorlinks=true,
    linkcolor=black,
    filecolor=magenta,      
    urlcolor=blue,
    citecolor=blue,
}

\lstdefinestyle{busquedas}{
    basicstyle=\ttfamily\footnotesize,
    breakatwhitespace=false,         
    breaklines=true,                 
    captionpos=t,                    
    keepspaces=true,                
    showspaces=false,                
    showstringspaces=false,
    showtabs=false
}

\newcolumntype{?}{!{\vrule width 2pt}}

\begin{document}

\title{XAI-Driven Spectral Analysis of Cough Sounds for Respiratory Disease Characterization}

\author{%
\textbf{Patricia Amado-Caballero}$^{1}$\thanks{Email: patricia.amado@uva.es} \and
\textbf{Luis M. San-José-Revuelta}$^{1,2}$\thanks{Email: lsanjose@tel.uva.es} \and
\textbf{María Dolores Aguilar-García}$^{3}$\thanks{Email: maguilarg@saludcastillayleon.es} \and
\textbf{José Ramón Garmendia-Leiza}$^{3}$\thanks{Email: jgarmendia@saludcastillayleon.es} \and
\textbf{Carlos Alberola-López}$^{1}$\thanks{Email: caralb@tel.uva.es} \and
\textbf{Pablo Casaseca-de-la-Higuera}$^{1}$\thanks{Corresponding author. Email: casaseca@lpi.tel.uva.es}%
}

\date{
  $^{1}$ ETSI de Telecomunicación, Universidad de Valladolid, Spain \\
  $^{2}$ Instituto de Neurociencias de Castilla y León (INCYL), Spain \\
  $^{3}$ Complejo Asistencial Universitario de Salamanca, SACYL, Spain
}

\maketitle

\begin{abstract}
This paper proposes an eXplainable Artificial Intelligence (XAI)-driven methodology to enhance the understanding of cough sound analysis for respiratory disease management. We employ occlusion maps to highlight relevant spectral regions in cough spectrograms processed by a Convolutional Neural Network (CNN). Subsequently, spectral analysis of spectrograms weighted by these occlusion maps reveals significant differences between disease groups, particularly in patients with COPD, where cough patterns appear more variable in the identified spectral regions of interest. This contrasts with the lack of significant differences observed when analyzing raw spectrograms. The proposed approach extracts and analyzes several spectral features, demonstrating the potential of XAI techniques to uncover disease-specific acoustic signatures and improve the diagnostic capabilities of cough sound analysis by providing more interpretable results.
\end{abstract}

\keywords{Deep Learning, eXplainable Artificial Intelligence, cough analysis, respiratory diseases, occlusion maps, spectral features.}



\section{Introduction}
\label{intro}

Respiratory diseases constitute a major cause of mortality worldwide~\cite{who17}, encompassing a broad spectrum of conditions. These range from acute infections, such as influenza and pneumonia, to chronic disorders like asthma and COPD, as well as severe illnesses like lung cancer. Chronic respiratory diseases are particularly concerning due to their long-term impact, often leading to sustained disability and dependence. Moreover, the COVID-19 pandemic represented a key moment in the understanding and management of these conditions. It highlighted the necessity of maintaining specific treatments at home when hospitalization was impractical and emphasized the importance of structured recovery protocols~\cite{Belli20}. As a matter of fact, chronic respiratory patients faced heightened vulnerability during the pandemic, with an increased risk of severe outcomes and mortality~\cite{who21}.

In light of the challenges described above, continuous monitoring of respiratory patients is essential to ensure timely and effective intervention, particularly during exacerbations or disease progression. A 2018 study by the European Commission highlighted the critical role of telemedicine in managing respiratory conditions~\cite{CE18}, stressing the need for further research in this domain. Although interest in this field has grown since the early 2010's~\cite{Audit11}, progress remains limited by the subjective nature of symptom assessment. Integrating objective measurements with real-time monitoring is key to facilitating early diagnosis and improving patient outcomes ~\cite{Pinnock13}.

As presented in Section \ref{sota}, numerous studies have explored the automatic detection and classification of cough events using traditional machine learning (ML) with hand-crafted audio features and deep learning (DL) techniques. Even though these methods, particularly DL models, have shown promising accuracy and sensitivity, diseases identification remains a challenge when large and diverse cough databases are not available. In addition, research establishing direct links between detected cough events and the underlying mechanisms of specific diseases remains scarce, likely due to the inherent ``black box'' nature of these models.

This paper aims to deepen this area by introducing a methodology based on eXplainable Artificial Intelligence (XAI) to identify disease-specific spectral patterns in cough sounds across different respiratory conditions. The proposed approach uses occlusion maps ~\cite{zeiler2014visualizing} to highlight relevant regions in cough spectrograms processed by a Convolutional Neural Network (CNN). Subsequent spectral analysis reveals significant differences between disease groups when features are extracted from spectrograms weighted by the occlusion maps, whereas no significant differences emerge when analyzing raw spectrograms alone.

The remainder of this paper is as follows: Section \ref{sota} reviews the state-of-the-art in automatic cough detection and diagnosis, covering relevant machine learning methods (both conventional and deep), as well as the application of XAI techniques to cough analysis. Section \ref{matmet} details the materials used in this study and the proposed methodology for identifying and interpreting cough-related features across different diseases. Results are presented and discussed in Section \ref{results}. Finally, Section \ref{conclusion} summarizes the key findings of the study.

\section{State of The Art}
\label{sota}

Cough is one of the most prevalent symptoms of respiratory diseases. Its detection has become a key focus in audio signal processing, especially with recent advances in Machine (Deep) Learning. Several algorithms have been used to develop efficient detectors that can identify cough events from different sources. Although some studies combine audio signals with other data, such as ECG or chest belt measurements~\cite{drugman2013objective}, the most recent research has focused on audio-based methods.

\subsection{ML-based Cough detection from hand-crafted features}

Most of the classical, ML-based detection methods~\cite{drugman2013objective,Birring2008,Vizel2010,swarnkar2013neural,sterling2014automated,drugman2014using,amrulloh2015automatic,KLCO201836} implement a pattern recognition module that classifies a set of hand-crafted time and/or spectral domain features extracted from the audio signals. Among the most common feature sets are those typically used in speech signal analysis such as MFCC (Mel Frequency Cepstral Coefficients)~\cite{tokuda94_icslp}, GTCC (GammaTone Cepstral Coefficients)~\cite{Liu2013}, which have been widely used. These spectral coefficients are based on a logarithmic analysis of the signal spectrum to emulate how the human auditory system works. Other employed feature sets such as Linear Prediction Cepstral Coefficients (LPCC)~\cite{Mammone96} or Spectral Subband Centroid Histograms (SSCH)~\cite{Gaj06} can also provide perceptually relevant information for cough detection. However, Monge et al.~\cite{monge2018robust} showed that these feature sets did not result in accurate detection of cough events in noisy conditions. Instead, they demonstrated that using local image moments over 2D time-frequency representations (spectrograms) of the audio signals solved the detection problem in heavily noise-contaminated signals~\cite{monge2018robust,casaseca2015effect,monge2016effect,hoyos2017efficient,hoyos2018efficient,monge2018audio}. Furthermore, they also demonstrated~\cite{Monge19} that considering temporal relationships between features extracted over short periods of time resulted in features of extended temporal duration that allowed for improved cough detection.

\subsection{Cough detection using DL}
\label{coughdetML}
Early proposals based on DL applied to cough detection performed automatic feature extraction using convolutional~\cite{liu2014cough,amoh2015deepcough,amoh2016deep,kadambi2018} and deep recurrent ~\cite{amoh2016deep} neural networks, although the starting point is usually the 2D time-frequency representation of audio signals, either directly or transformed into a space of perceptual coefficients such as MFCCs. The results of these preliminary works were promising and, in many cases, outperformed classical methods in which the feature set was predefined. However, these techniques present a generalization problem in part due to the need for large training databases. 

Since 2020, there has been an explosion in the use of DL-based methods to solve the problem of cough detection and analysis. This has been primarily due to the need for useful diagnostic and screening tools for COVID-19. Most of the cough detection proposals~\cite{kvapilova2020continuous,li2020eeg} employ standalone convolutional neural networks (CNNs), while others~\cite{you2022automatic} integrate CNNs with recurrent architectures, such as Long Short-Term Memory (LSTM) layers, to capture the temporal structure of cough events. 

Today, DL-based approaches significantly outperform traditional ML methods in cough detection, effectively addressing the generalization issues of earlier models. This progress is largely attributed to the availability of large, diverse cough sound databases, crowd-sourced during the pandemic. Relevant datasets~\cite{Uni21, bhattacharya2023coswara, Kha21} include voluntary forced cough recordings from COVID-19 patients, individuals with other respiratory conditions, and non-pathological cases such as smokers.

\subsection{Cough analysis for diagnosis support}

Beyond the importance of cough detection in tracking and monitoring exacerbations in respiratory diseases, there is growing interest in analyzing and characterizing cough sounds for diagnostic and screening purposes. Understanding the acoustic features of coughs could provide valuable insights for early detection and differentiation of respiratory conditions, ultimately improving patient outcomes. 

Some proposals based on traditional machine learning have been shown to be applicable to the analysis and characterisation of cough sounds for diagnostic purposes. This fact reveals the potential of cough to study the underlying mechanisms in diseases that present it as a symptom. Swarnkar et al. \cite{swarnkar2013automatic} used a logistic regressor fed with time and frequency audio features to differentiate between wet and dry coughs in paediatric patients. A similar approach was followed in \cite{abeyratne2013cough} to diagnose childhood pneumonia. This feature set was augmented with wavelet features \cite{kosasih2014wavelet} to improve specificity in the diagnosis. In ~\cite{amrulloh2015cough}, neural networks were used to differentiate pneumonia from asthma patients in a small patient group. Sharan et \textit{al.}~\cite{sharan2018automatic} used SVMs and Logistic regressors fed with MFCCs and chlocheagram image features (CIF) to diagnose croup in children.

The COVID-19 pandemic not only represented a health challenge, also prompted the need for an improvement in the diagnosis of respiratory diseases. As reported in section \ref{coughdetML}, the use of DL-based methods to solve the problem of cough detection and analysis for diagnostic purposes has increased significantly since 2020. Numerous studies have explored the use of cough sounds to distinguish COVID-19 from other diseases. A comprehensive review of research conducted up to 2022 can be found in ~\cite{Ghrabli22}, with key studies summarized in the following paragraphs.

Laguarta et {\it al}~\cite{laguarta2020covid} used CNN for diagnosis with MFCC as inputs, achieving a sensitivity value of 98.5\%. Imran et {\it al.}~\cite{imran2020ai4covid} combined CNNs with Support Vector Machines (SVM), achieving an accuracy of  92.64\%. Tena {\it al.}~\cite{tena2022automated} used YAMNET, a deep neural network intended for audio classification, to identify cough sounds, whose spectrograms were further classified using time and spectral features for COVID detection with accuracies close to 90\%. Pahar et al.~\cite{pahar2020covid} developed a DL-based COVID-19 cough classifier capable of distinguishing COVID-19 positive coughs from both COVID-19 negative and healthy coughs recorded on a smartphone. Using a ResNet50 classifier, they achieved an AUC of 98\% when differentiating COVID-19 positive from healthy coughs, while an LSTM classifier performed best in distinguishing COVID-19 positive from COVID-19 negative coughs, reaching an AUC of 94\%. Similarly, Brown et al. \cite{brown2020exploring} explored the use of cough and breathing sounds to assess how well COVID-19 cases could be differentiated from asthma or healthy controls. Their results showed that even a simple binary machine learning classifier could reliably distinguish healthy from COVID-19 sounds, with models achieving an AUC above 80\% across all tasks.

Other DL-based studies have reported lower performance. For example, Vrindavanam et al. \cite{vrindavanam2021machine} achieved an accuracy of 70.6\% using a convolutional neural network (CNN). Feng et al. \cite{Feng2021DeeplearningBA} obtained 90\% accuracy with RNN trained on the Coswara dataset~\cite{bhattacharya2023coswara}. However, when combining Coswara with Virufy~\cite{Kha21}, accuracy dropped to 80, underscoring the challenge of maintaining robust performance across multiple datasets. More advanced architectures, including Audio Spectrogram Transformers (AST)~\cite{gong2021ast}, have also been used to differentiate dry and wet cough from other sounds~\cite{habashy2022cough}, achieveing 81\% F1-score.

More recent studies (post-2024) have implemented advanced approaches with notable success. Hussain et al.~\cite{hussain2024cough2covid19} employed a multi-layer ensemble deep learning (MLEDL) model to analyze cough audio features such as MFCC, spectrograms, and chromagrams, achieving a 98\% accuracy rate for COVID-19 detection across public datasets. Similarly, Islam et al.~\cite{islam2025robustcovid19detectioncough} used deep neural decision trees and random forest models trained on similar features, reporting accuracies of up to 97\% across diverse datasets. Finally, foundation models,based on transformers trained on large volumes of unlabeled data, have recently been developed for respiratory signal and cough analysis \cite{baur2024hear}. Laska et al.\cite{laska2024zero} achieved high performance in key tasks, including cough/breathing/speech discrimination (100\%), cougher verification (96.9\%), cougher identification (84.4\%), and wet/dry cough classification (93.8\%).

\subsection{Addressing Research Gaps in Cough Analysis: Using XAI for Disease Differentiation}

ML-driven cough detection and analysis have rapidly emerged as a dynamic and intensely researched domain. Early efforts were dominated by classical feature-based ML models, which heavily rely on handcrafted features extracted from audio signals. However, these systems often face significant challenges in identifying and leveraging informative features. While successful feature selection can yield valuable insights into the underlying causes of diseases, this process is frequently hindered by two key issues: either such features cannot be identified, or they are so complex that they become difficult to interpret. 

Most of the classical ML approaches methods have been surpassed by modern DL methods. However, despite their effectiveness, deep neural networks inherently function as ``black-box'' models, offering little to no insight into the underlying factors driving their predictions without the application of specific eXplainable AI (XAI) techniques. 

There is thus potential in XAI methods in providing deeper insight into the specifics of cough for different diseases. Considering classical ML, in~\cite{sanim2023identification,wullenweber2022coughlime} Local Interpretable Model-Agnostic Explanations (LIME)~\cite{ribeiro2016should} were employed to identify the relevance of different audio features in differentiating COVID-19 from other respiratory diseases. As for DL approaches, in~\cite{Shen2024}, Gradient-weighted Class Activation Mapping (Grad-CAM)~\cite{selvaraju2017grad} was employed  to show the importance of CNN-extracted features in differentiating COVID-19, asthma, pneumonia and healthy subjects. Similarly, in~\cite{amado2024}, occlusion maps~\cite{zeiler2014visualizing} were employed to identify the spectrogram regions that contributed most significantly to CNN-based cough detection. By fitting a Gaussian Mixture Model (GMM) to cough spectrograms weighted by occlusion maps, some regions were found to differ markedly energy-wise between diseases. However, two main limitations are observed in using such GMMs to account for disease-specific differences:
\begin{enumerate}
    \item The employed bimodal GMM did not provide the best fit for certain cough spectrograms, leading to potential inaccuracies that could misguide interpretation.
    \item The explainability obtained with the model parameters was only reliant into differences in the time nature of the cough event. This made the method heavily dependent on the precise cough starting point. As a result, the cough detection method would need to go beyond identifying audio clips containing coughs. It must also accurately delineate the event by locating its exact start and end timestamps.
\end{enumerate}

In this paper we employ XAI methodologies to identify key spectral information in cough spectrograms and extract relevant parameters from XAI-enhanced representations. We demonstrate that these XAI-driven features can reveal disease-specific differences valuable for diagnosis that remained hidden when derived from non-enhanced spectrograms. Since parameters are directly extracted from the enhanced spectrograms, our approach eliminates reliance on the GMM model goodness-of-fit, unlike the approach in \cite{amado2024}, with which we compare our results. Additionally, the spectral nature of the extracted features allows the method to operate without requiring precise delineation of cough events.

\section{Materials and methods}
\label{matmet}

\subsection{Materials}
\label{mat}
We conducted an observational study tracking cough patterns over a 24-hour period in real-world conditions, analyzing approximately 15,000 single cough events. To achieve this, we prospectively recorded full-day audio samples from ambulatory patients within the Palencia Health Area (Spain). Data collection was performed using a Sony Xperia Z2 Android smartphone, capturing audio in 16-bit WAV format at a sampling rate of 44.1 kHz. The group included 20 patients aged between 23 and 87 years (7 women, 10men) with the following respiratory pathologies: Acute respiratory disease (ARD, 3), pneumonia (3), COPD (6), lung cancer
(2), and others such as asthma, bronchiectasis or sarcoidosis (remaining patients). The patients were divided into 6 comparison
groups marked by the type of pathologies diagnosed. As in \cite{amado2024}, the following sets were defined for comparison (see table \ref{tab:grupos}):
\begin{itemize}
    \item G1: Chronic vs. Non-chronic patients.
    \item G2: COPD patients vs. other diseases.
    \item G3: COPD patients vs. other diseases excluding cancer.
    \item G4: COPD patients vs. ARD and pneumonia patients.
    \item G5: COPD patients vs. other chronic diseases.
    \item G6: COPD patients vs. lung cancer patients
\end{itemize}

\begin{table}[!ht]
\caption{Summary of the patient database. Pathologies on the left hand side are chronic; those on the right hand side are referred to as non-chronic.}
\label{tab:grupos}
\centering
\begin{tabular}{|c|l?c|l|}
\hline
\multicolumn{2}{|c?}{\textbf{CHRONIC}}  & \multicolumn{2}{c|}{\textbf{NON-CHRONIC}}  \\ \cline{1-2}
\hline\hline 
COPD            & 6 & \multicolumn{1}{c|}{\multirow{2}{*}{Pneumonia}} & \multirow{2}{*}{3} \\ \cline{1-2}
Lung Cancer    & 2 & \multicolumn{1}{l|}{}                           &                    \\ \hline
Asthma         & 1 & \multirow{3}{*}{ARD}                            & \multirow{3}{*}{3} \\ \cline{1-2}
Sarcoidosis     & 1 &                                                 &                    \\ \cline{1-2}
Bronchiectasis & 1 &                                                 &                    \\ \hline
\end{tabular}
\end{table}

\subsection{Methods}
\label{sec:Methods}

\subsubsection{Audio signal preprocessing}

The audio signal undergoes $5\times$ down-sampling, followed by windowing to segment the data effectively. Next, spectrograms are computed to generate 2D time-frequency representations, which serve as input to a CNN that classifies each window as either cough or non-cough. The detailed procedure is outlined below: 
\begin{itemize}
\item The power spectral density (PSD) is computed for 10 ms segments using a Hanning window with no overlap.
\item These segments are then concatenated into 1-second intervals, producing 45×100 spectrograms.
\item Finally, the spectrograms undergo logarithmic normalization, transforming them into input images optimized for the CNN.
\end{itemize}

\subsubsection{Cough Window Identification}

We employed the CNN in~\cite{amado2024} for detection and identification of meaningful cough spectrograms. The architecture begins with a convolutional layer containing 32 filters, each with $2\times2$ kernels, activated by ReLU, followed by a $2\times2$ Max-Pooling layer to reduce dimensionality and a dropout layer to prevent overfitting. This structure is then repeated with the number of filters doubled at each stage. The final sequence before the output consists of a convolutional layer with 128 filters, another dropout layer, a subsequent convolutional layer with 256 filters, and a concluding Max-Pooling layer. The extracted features are then reshaped to match the output architecture, which includes two Fully-Connected layers: the first with 512 neurons and ReLU activation, and the second with two neurons employing a softmax function for classification.

\begin{figure*}[!htb]
\centering
\includegraphics[width=0.75\textwidth]{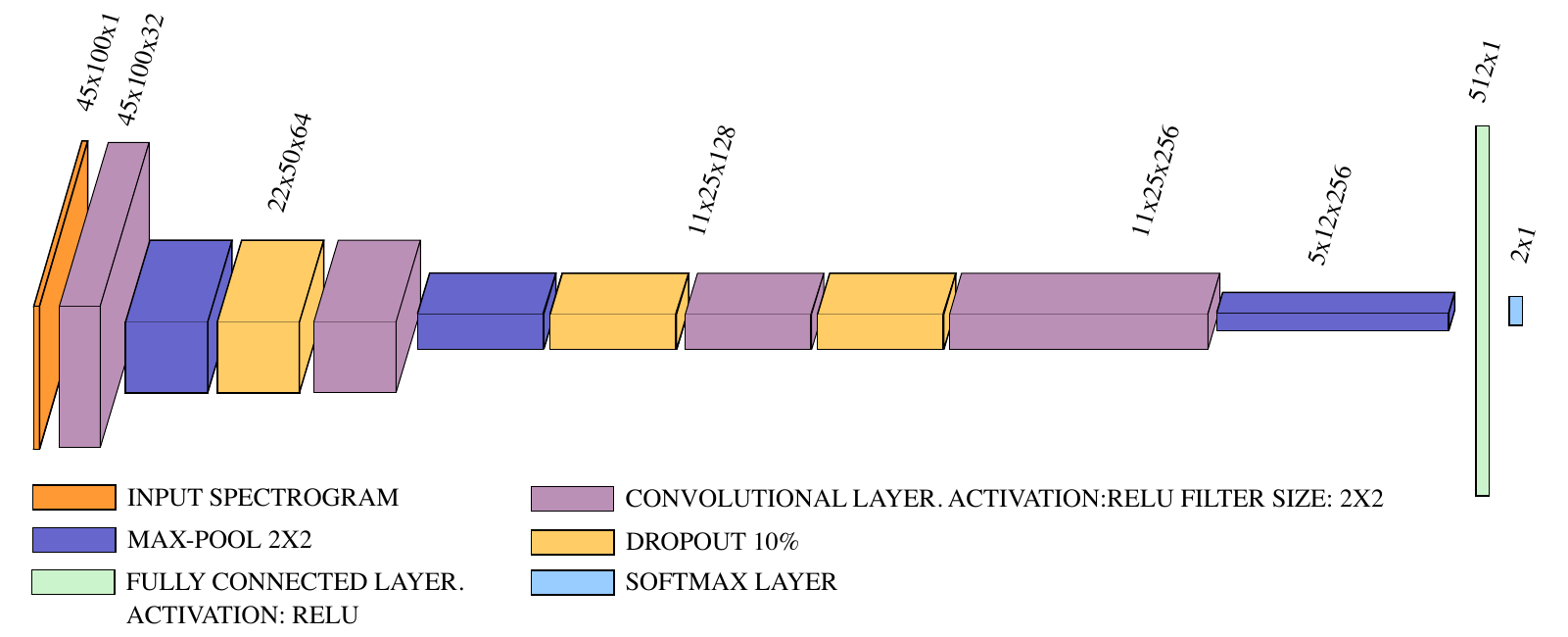}
\captionsetup{justification=centering}
\caption{CNN architecture for cough detection.}
\label{figCNN}
\end{figure*}

The training process was carried out using the AdaMax optimizer ($\alpha=0.002$), batch size$=128$ and 50 epochs. To avoid overfitting, we employed a validation set comprising $20\%$ of the training dataset, which itself constituted $80\%$ of the entire audio clips collection. Train and test sets were organized in 5-folds to ensure that coughs belonging to patients in the training set were never in its corresponding test set in the fold and all the patients were at least once in a test set. Figure \ref{figCNN} illustrates a block diagram of the employed CNN.

\subsubsection{XAI-driven cough analysis} 

\begin{figure*}[htbp]
\centerline{\includegraphics[width=4.5cm]{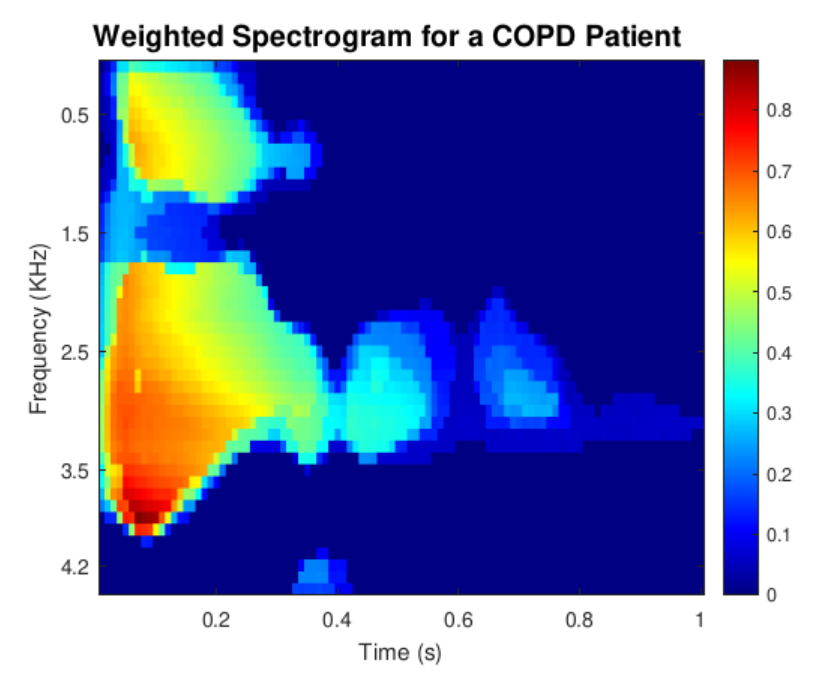}
\includegraphics[width=4.5cm]{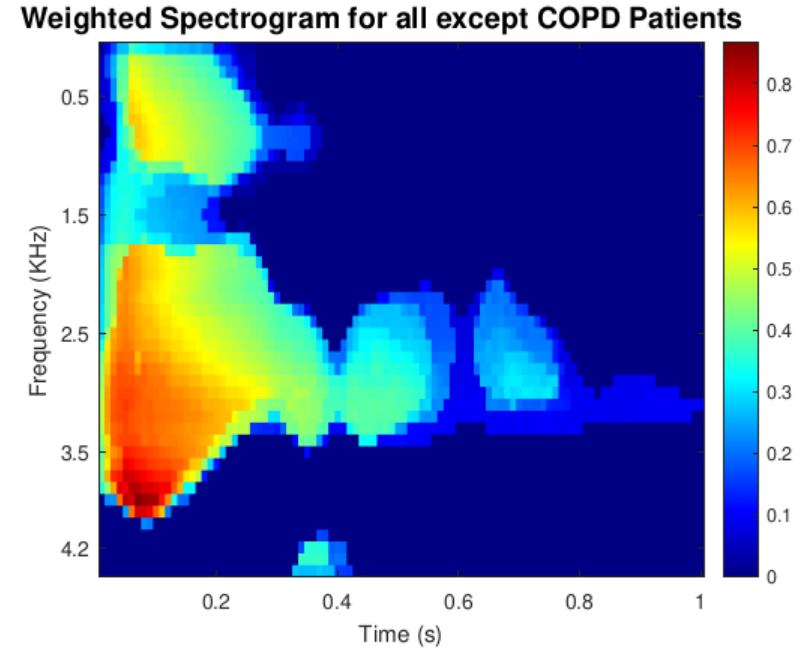}
\includegraphics[width=4.5cm]{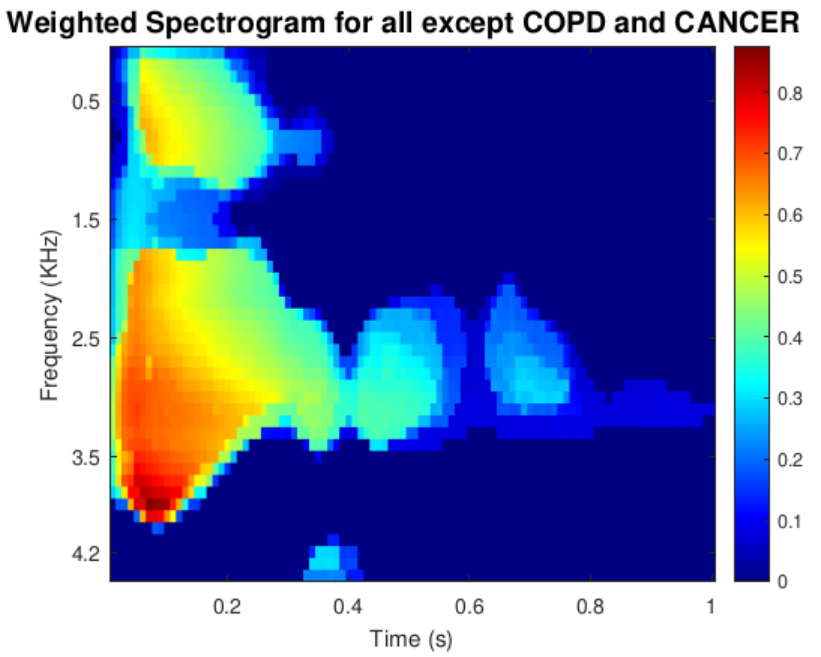}
\includegraphics[width=4.5cm]{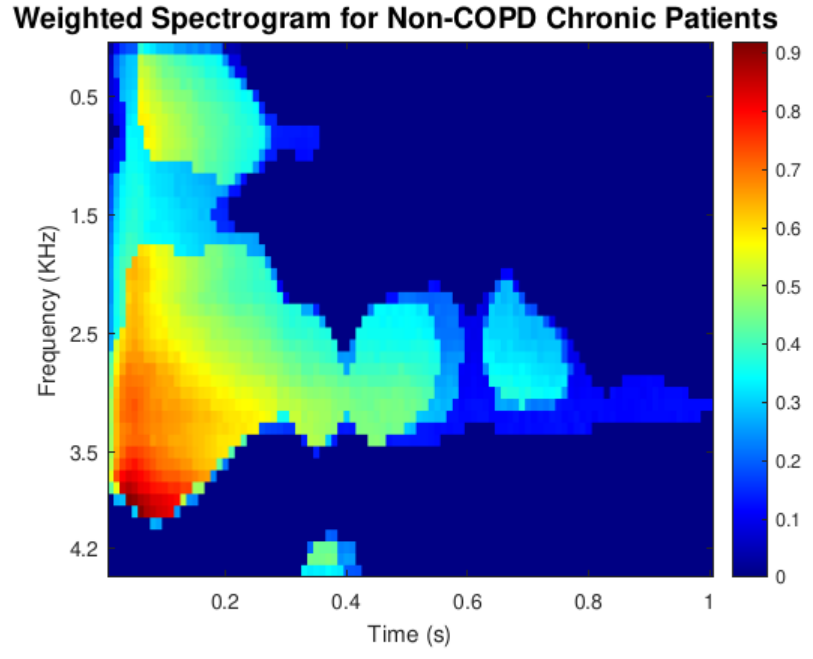}}
\centerline{\includegraphics[width=4.5cm]{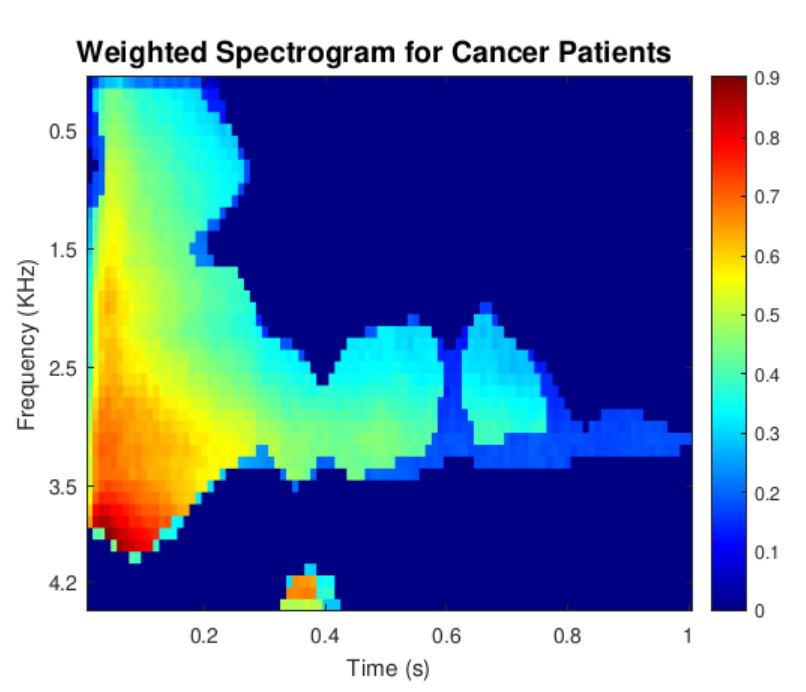}
\includegraphics[width=4.5cm]{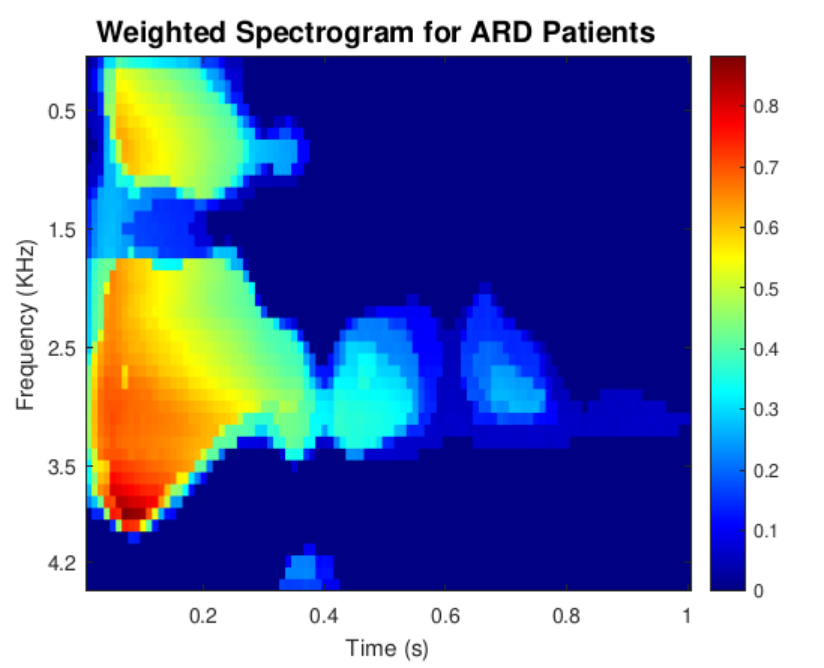}
\includegraphics[width=4.5cm]{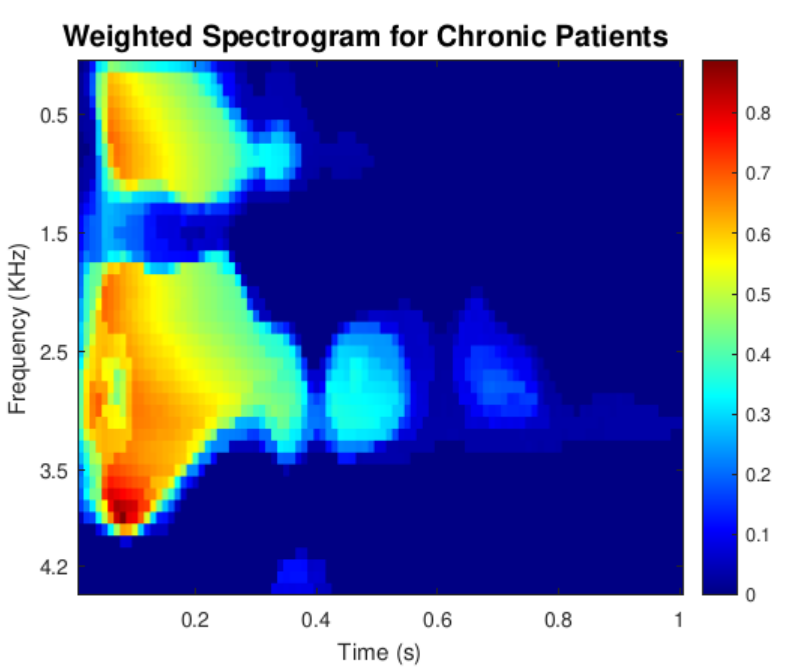}
\includegraphics[width=4.5cm]{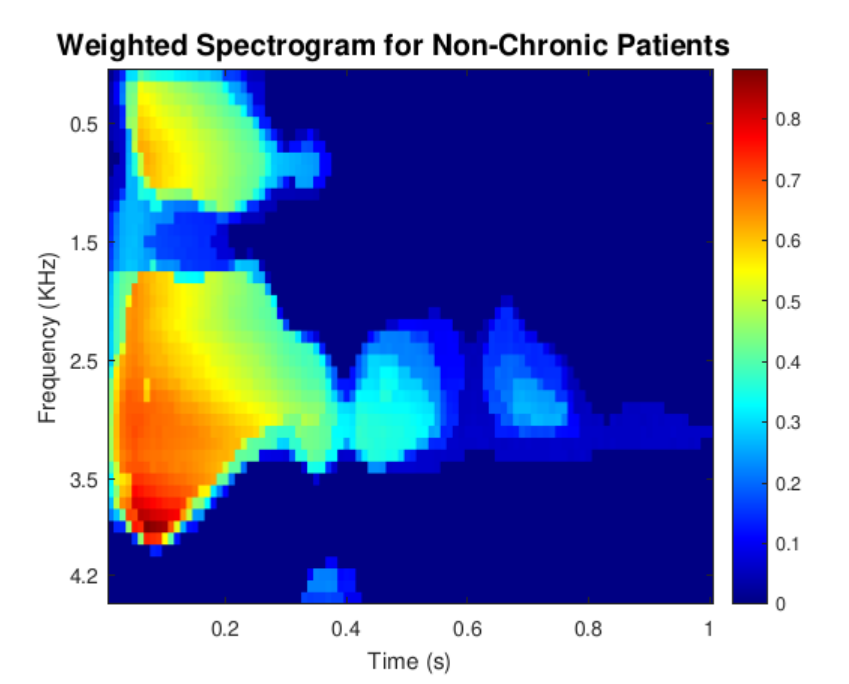}}
\caption{Median Weighted Spectrogram for the pathology sets}
\label{figweightedspectrograms}
\end{figure*}

Once the CNN was trained, its behavior over the test set was analyzed to identify relevant areas for cough analysis within the spectrograms. This analysis was based on XAI techniques as follows: 

\begin{itemize}
    \item \textit{Occlusion Maps}: We computed occlusion maps ~\cite{zeiler2014visualizing} on spectrograms classified as cough by the CNN with confidence levels greater than $90\%$. Generating the occlusion maps involves applying a mask to the input spectrogram, obscuring part of its information from the CNN. The class probability produced by the CNN with this modified input helps evaluate the significance of the occluded region for classification (a higher probability meaning lower importance). This is repeated over the whole spectrogram. Once all the maps were created, they were resized to have the same dimensions as the original input. Then, they were normalized and pixel-averaged. After this process, the cough pattern of each patient was represented with a unique occlusion map.
    \item \textit{Weighted Spectrograms}: Once the occlusion maps were generated, each patient’s pixel-averaged spectrogram was weighted by its corresponding map. Specifically, we pixelwise multiplied the pixel-averaged spectrogram and the thresholded map, so we either multiplied by zero (pixel map under threshold) or by the value of the map (pixel map exceeding threshold --Th.--):
    \begin{equation}
  \mathcal{S}[k,n] = \mathcal{S}_0[k,n] \odot \left [\mathcal{M}[k,n]>\text{Th}\right ]
\end{equation}
where $\mathcal{S}[k,n]$ denotes the weighted spectrogram, $\mathcal{S}_0[k,n]$ the original spectrogram, $\mathcal{M}[k,n]$ the occlusion map, and $\odot$ the Hadamard (element-wise) product.

    This approach ensured that only the most relevant regions 
    were subjected to further analysis. The outcome is an enhanced spectrogram that selectively amplifies the highlighted regions, improving the interpretability of the features. Figure \ref{figweightedspectrograms} shows examples of the median of the weighted spectrograms for each of the studied groups.
\end{itemize}

\begin{figure*}[!htb]
\centering
\includegraphics[width=1\textwidth]{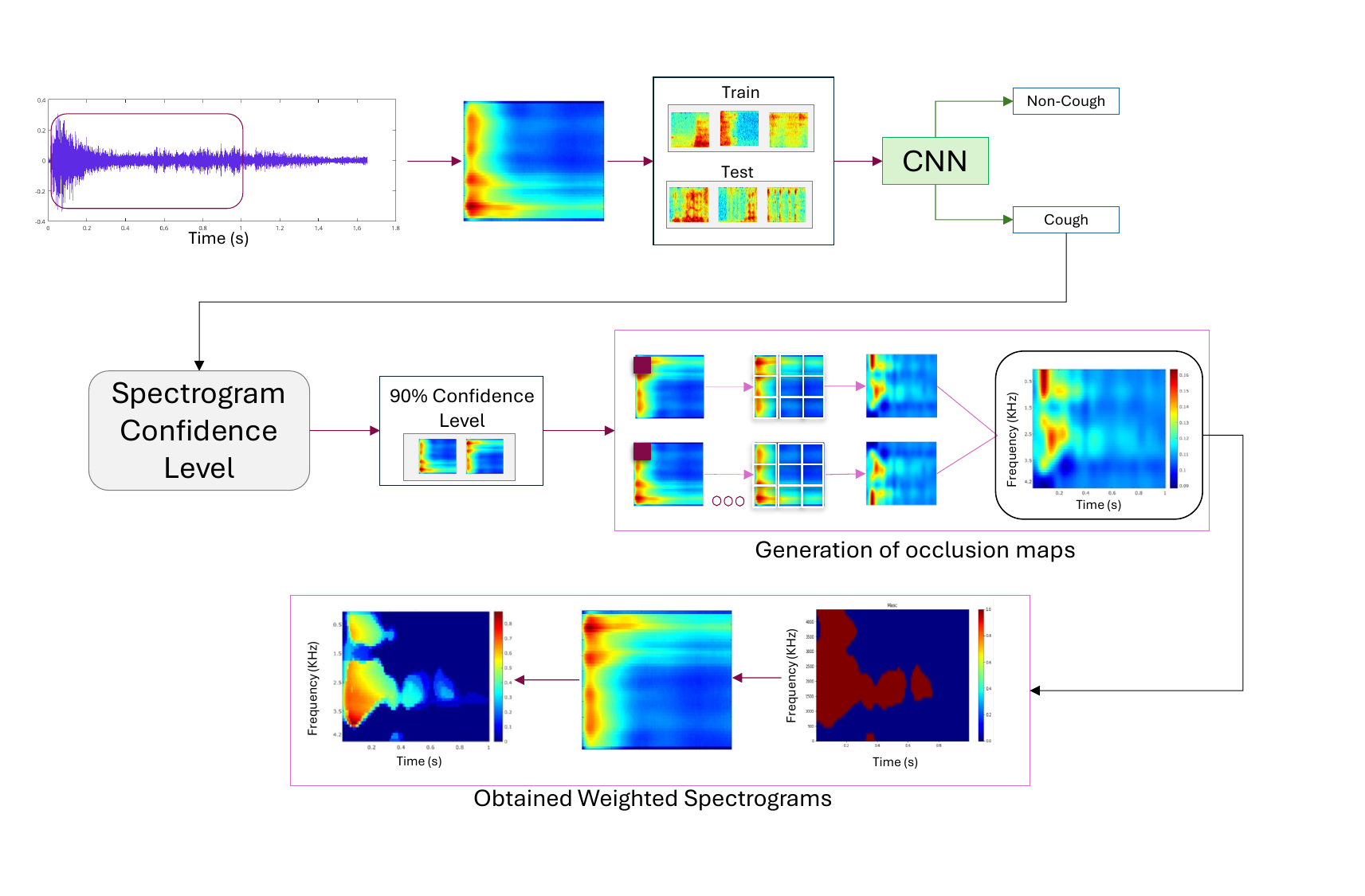}
\captionsetup{justification=centering}
\caption{Overview of the classification and XAI-driven cough analysis.}
\label{figMet}
\end{figure*}

\subsubsection{Spectral features obtained from the Weighted Spectrograms}  

Spectral features are typically used in audio analysis~\cite{ramalingman2006}. Some of these features have previously been used to identify cough patterns~\cite{Monge19}, mainly due to their low computational complexity over the spectrum. Beyond their efficiency, these features offer a meaningful physical interpretation without requiring precise identification of the cough event’s onset, as they inherently capture frequency-based characteristics. Thus, we obtained these features from the weighted spectrograms to find patterns inherent to the different diseases.

The calculated features are described in the following paragraphs. We refer to the discrete (weighted) one-sided spectrogram as $\mathcal{S}[k,n]$, where $k=0,\ldots,K$ denotes de frequency index ($K=44$), being the discrete frequencies $f[k]=k\cdot f_s/(2\cdot K+1)$, with $f_s=44.1/5=8.82$ kHz. On the other hand, $n=0,\ldots,N-1$ stands for the time index ($N=100$).

The first extracted feature is the \textbf{Relative AC power}, obtained as the ratio between the AC and total power:  
\begin{equation}
AC= \frac{1}{N}\sum_{n=0}^{N-1}\left[\frac{\sum\limits_{k=1}^{K}\mathcal{S}[k,n]}{\sum\limits_{k=0}^{K}\mathcal{S}[k,n]}\right]   
\label{eq:RP}
\end{equation}

The \textbf{Spectral bandwith} was calculated as a measure of the spread of the spectral distribution as
\begin{equation}
SpBW=\frac{1}{N}\displaystyle\sum\limits_{n=0}^{N-1}\left[
\frac{\sum\limits_{k=0}^{K}(f[k]-SpC[n])^2\cdot\mathcal{S}[k,n]}{\sum\limits_{k=0}^{K}\mathcal{S}[k,n]}
\right]
\label{eq:Specband}
\end{equation}
where $SpC[n]$ is the \textbf{Spectral centroid}, representing the spectral center of gravity as:
\begin{equation}
SpC[n]= \frac{\sum\limits_{k=0}^{K}f[k]\cdot\mathcal{S}[k,n]}{\sum\limits_{k=0}^{K}\mathcal{S}[k,n]}\label{eq:SpecCent}
\end{equation}

The \textbf{Spectral Crest Factor} was also computed as a feature used to detect the dominant frequency of the spectrum. 

\begin{equation}
SpCF= \frac{1}{N}\displaystyle\sum_{n=0}^{N-1}\left[
\frac{\max\limits_k(\mathcal{S}[k,n])}{C\cdot\sum\limits_{k=0}^{K}\mathcal{S}[k,n]}
\right]
\label{eq:SpecCrest}
\end{equation}
with $C= 1/(\max(f[k])-\min(f[k])+1)$.

The \textbf{Spectral flatness} was obtained as a measure quantifying the degree of flatness of the spectrum. A high value represents a white-noise-like distribution~\cite{ramalingman2006}.
\begin{equation}
SpF=\frac{1}{N}\displaystyle\sum\limits_{n=0}^{N-1}\left[
\frac{\exp\left(\frac{1}{K+1}\sum\limits_{k=0}^{K}\log(\mathcal{S}[k,n])\right)}{\frac{1}{K+1}\sum\limits_{k=0}^{K}\mathcal{S}[k,n]} 
\right]
\label{eq:SpecFlat}
\end{equation}

To quantify the spectral variation between two consecutive time stamps, we computed the \textbf{Spectral flux}~\cite{giannakopoulos2014introduction}:
\begin{equation}
SpFX= \frac{1}{N-1}\displaystyle\sum_{n=1}^{N-1}\left[
\sum_{k=0}^{K}(\mathcal{S}[k,n]-\mathcal{S}[k,n-1])
\right]
\label{eq:SpecFlux}
\end{equation}

The \textbf{Spectral Renyi entropy} was also computed as a generalized measure of uncertainty or randomness.
\begin{equation}
SpRE= \frac{1}{N}\displaystyle\sum_{n=0}^{N-1}\left[
\frac{1}{1-q}\cdot\log\left(\sum_{k=0}^{K}{\mathcal{S}[k,n]}\right) ^q 
\right]
\label{eq:SpecRenyiEnt}
\end{equation}
where $q=4$ was used.

Finally, we defined the \textbf{Spectral roll-off} to account for the frequencies representing the 85th percentile of the total power~\cite{giannakopoulos2014introduction}:
\begin{equation}
    SpR=\frac{1}{N}\displaystyle\sum_{n=0}^{N-1}f[k_{85}[n]]
 \label{eq:SpecRollOff}
\end{equation}
with $k_{85}[n]$ the minimum value of $k$ for which:
\begin{equation}
\sum_{k=0}^{k_{85}[n]}\mathcal{S}[n,k] \geq 0.85\cdot\sum_{k=0}^{K}\mathcal{S}[n,k]
\end{equation}

\subsubsection{Testing for group differences}

To identify statistically significant differences between groups in each set (G1--G6) defined in section \ref{mat}, the spectral features extracted for each group were analyzed using hypothesis testing. Initially, a Gaussianity test was performed. If both groups satisfied Gaussianity, an unpaired Student’s t-test was applied. Otherwise, the Mann-Whitney U-test was used. After conducting the tests for all features, boxplots were generated for cases where significant differences were identified (p-value < 0.05).

\section{Results and Discussion}
\label{results}
\begin{table*}[!htb]

\caption{Separability results obtained for groups G1--G6 with the proposed method. $p$-values are obtained from hypothesis testing over spectral features extracted from XAI-driven \textit{Weighted Spectrograms}. Statistical significance is considered for $p<0.05$ and highlighted using green, boldfaced font.}
\label{tab:featuresTotal1}
\resizebox{\textwidth}{!} {
\begin{tiny}
\begin{tabular}{|c|ccccccc|}
\noalign{\hrule height 1pt }
\cline{2-8}
\multicolumn{1}{|l|}{Th.}                                        & \multicolumn{1}{c|}{AC}                             & \multicolumn{1}{c|}{\textbf{$SpBW$}}                       & \multicolumn{1}{c|}{\textbf{$SpCF$}}                                           & \multicolumn{1}{c|}{\textbf{$SpF$}}                       & 
\multicolumn{1}{c|}{\textbf{$SpFx$}}                       & 
\multicolumn{1}{c|}{\textbf{$SpRE$}}                       & \textbf{$SpR$}                         \\ \cline{2-8}
\hline \hline 
\multicolumn{8}{|c|}{\textbf{G1}} \\ \hline
{\textbf{0.5}} & \multicolumn{1}{c}{0.7396}  
& \multicolumn{1}{c}{0.9623}
& \multicolumn{1}{c}{{\color[HTML]{036400} \textbf{0.0330}}}        
& \multicolumn{1}{c}{0.6691}    
& \multicolumn{1}{c}{0.8125}
& \multicolumn{1}{c}{0.3148}&0.4747   \\ \hline
{\textbf{0.6}} & \multicolumn{1}{c}{0.4173}  
& \multicolumn{1}{c}{{0.1331}}
& \multicolumn{1}{c}{0.4747}       
& \multicolumn{1}{c}{0.2295}    
& \multicolumn{1}{c}{0.7396}
& \multicolumn{1}{c}{0.6691}& 0.3148        \\ 
\hline
{\textbf{0.7}} & \multicolumn{1}{c}{0.1088}  
& \multicolumn{1}{c}{0.0702}
& \multicolumn{1}{c}{0.2295}   
& \multicolumn{1}{c}{0.0553}    
& \multicolumn{1}{c}{0.5362}
& \multicolumn{1}{c}{0.0702}& 0.0878       \\ \hline
{\textbf{0.8}} & \multicolumn{1}{c}{0.8868}  
& \multicolumn{1}{c}{0.4747}
& \multicolumn{1}{c}{0.2698}      
& \multicolumn{1}{c}{0.8125}    
& \multicolumn{1}{c}{0.7396}
& \multicolumn{1}{c}{0.7396}& 0.4173        \\ \hline
{\textbf{0.9}}  & \multicolumn{1}{c}{1} 
& \multicolumn{1}{c}{0.7396}        
& \multicolumn{1}{c}{0.4173}    
& \multicolumn{1}{c}{0.6691} & \multicolumn{1}{c}{0.4747}
& \multicolumn{1}{c}{0.7396}& 0.7212  \\ \hline
\hline
 \multicolumn{8}{|c|}{\textbf{G2}} \\
 \hline
 {\textbf{0.5}} & \multicolumn{1}{c}{0.1215}  
& \multicolumn{1}{c}{0.3502}
& \multicolumn{1}{c}{0.0616}    
& \multicolumn{1}{c}{0.5249}    
& \multicolumn{1}{c}{0.5249}
& \multicolumn{1}{c}{0.3502}& {0.0616}       \\ \hline
{\textbf{0.6}} & \multicolumn{1}{c}{0.6605}  
& \multicolumn{1}{c}{0.1490}
& \multicolumn{1}{c}{0.9612}        
& \multicolumn{1}{c}{\color[HTML]{036400} \textbf{0.0011}}
& \multicolumn{1}{c}{0.7325}  
& \multicolumn{1}{c}{{\color[HTML]{036400} \textbf{0.0145}}} & {{\color[HTML]{036400} \textbf{0.0202}}}       \\ 
\hline
{\textbf{0.7}} & \multicolumn{1}{c}{{\color[HTML]{036400} \textbf{0.0273}}} 
& \multicolumn{1}{c}{{\color[HTML]{036400} \textbf{0.0202}}}
& \multicolumn{1}{c}{{\color[HTML]{036400} \textbf{0.0103}}}      
& \multicolumn{1}{c}{{\color[HTML]{036400} \textbf{0.0202}}}  
& \multicolumn{1}{c}{0.7325}
& \multicolumn{1}{c}{{\color[HTML]{036400} \textbf{0.0145}}}& {{\color[HTML]{036400} \textbf{0.0103}}}       \\ \hline
{\textbf{0.8}} & \multicolumn{1}{c}{0.2561}  
& \multicolumn{1}{c}{0.2561}
& \multicolumn{1}{c}{1}      
& \multicolumn{1}{c}{0.3011}    
& \multicolumn{1}{c}{0.3011}
& \multicolumn{1}{c}{0.3011}& {{\color[HTML]{036400} \textbf{0.0447}}}        \\ \hline
{\textbf{0.9}}  & \multicolumn{1}{c}{0.4623}
& \multicolumn{1}{c}{0.5908}        
& \multicolumn{1}{c}{0.1490}    
& \multicolumn{1}{c}{0.4623} & \multicolumn{1}{c}{0.3011}
& \multicolumn{1}{c}{0.5249}& {{\color[HTML]{036400} \textbf{0.0452}}}      \\ \hline
\hline
\multicolumn{8}{|c|}{\textbf{G3}} \\ \hline
{\textbf{0.5}} & \multicolumn{1}{c}{0.1447}  
& \multicolumn{1}{c}{0.3884}
& \multicolumn{1}{c}{0.0663}       
& \multicolumn{1}{c}{0.8639}    
& \multicolumn{1}{c}{0.6889}
& \multicolumn{1}{c}{0.5287}& {0.0879}       \\ \hline
{\textbf{0.6}} & \multicolumn{1}{c}{0.6070}  
& \multicolumn{1}{c}{0.2238}
& \multicolumn{1}{c}{0.7756}        
& \multicolumn{1}{c}{{\color[HTML]{036400} \textbf{0.0028}}} 
& \multicolumn{1}{c}{0.8639}
& \multicolumn{1}{c}{{0.0360}}& {{\color[HTML]{036400} \textbf{0.0360}}}     \\ 
\hline
{\textbf{0.7}} & \multicolumn{1}{c}{{\color[HTML]{036400} \textbf{0.0360}}}
& \multicolumn{1}{c}{{\color[HTML]{036400} \textbf{0.0256}}}
& \multicolumn{1}{c}{{\color[HTML]{036400} \textbf{0.0256}}}       
& \multicolumn{1}{c}{{\color[HTML]{036400} \textbf{0.0256}}}  
& \multicolumn{1}{c}{0.7756}
& \multicolumn{1}{c}{{\color[HTML]{036400} \textbf{0.0176}}}& {{\color[HTML]{036400} \textbf{0.0120}}}       \\ \hline
{\textbf{0.8}} & \multicolumn{1}{c}{0.3884}  
& \multicolumn{1}{c}{0.1810}
& \multicolumn{1}{c}{1}      
& \multicolumn{1}{c}{0.3277}    
& \multicolumn{1}{c}{0.3884}
& \multicolumn{1}{c}{0.3277}& 0.0663        \\ \hline
{\textbf{0.9}}  & \multicolumn{1}{c}{0.6070} 
& \multicolumn{1}{c}{0.6889}        
& \multicolumn{1}{c}{0.2238}    
& \multicolumn{1}{c}{0.4559} & \multicolumn{1}{c}{0.2721}
& \multicolumn{1}{c}{0.6070}& 0.0627  \\ \hline
\hline
\multicolumn{8}{|c|}{\textbf{G4}}   \\ \hline
{\textbf{0.5}} & \multicolumn{1}{c}{0.1797}  
& \multicolumn{1}{c}{0.4848}
& \multicolumn{1}{c}{0.0649}        
& \multicolumn{1}{c}{0.4458}    
& \multicolumn{1}{c}{0.9372}
& \multicolumn{1}{c}{0.9372}& 0.0931      \\ \hline
{\textbf{0.6}} & \multicolumn{1}{c}{0.5887}  
& \multicolumn{1}{c}{0.2403}
& \multicolumn{1}{c}{0.8182}    
& \multicolumn{1}{c}{{\color[HTML]{036400} \textbf{0.0152}}}     
& \multicolumn{1}{c}{0.8182}
& \multicolumn{1}{c}{0.0649}& {{\color[HTML]{036400} \textbf{0.0411}}}  \\ 
\hline
{\textbf{0.7}} & \multicolumn{1}{c}{{\color[HTML]{036400} \textbf{0.0260}}}  
& \multicolumn{1}{c}{{\color[HTML]{036400} \textbf{0.0152}}}  
& \multicolumn{1}{c}{{\color[HTML]{036400} \textbf{0.0087}}}       
& \multicolumn{1}{c}{{\color[HTML]{036400} \textbf{0.0260}}}     
& \multicolumn{1}{c}{0.3095}
& \multicolumn{1}{c}{{\color[HTML]{036400} \textbf{0.0152}}}  & {{\color[HTML]{036400} \textbf{0.0152}}}        \\ \hline
{\textbf{0.8}} & \multicolumn{1}{c}{0.4848}  
& \multicolumn{1}{c}{0.1797}
& \multicolumn{1}{c}{1}        
& \multicolumn{1}{c}{0.3095}    
& \multicolumn{1}{c}{0.3939}
& \multicolumn{1}{c}{0.3095}& 0.1797        \\ \hline
{\textbf{0.9}}  & \multicolumn{1}{c}{0.6991}
& \multicolumn{1}{c}{0.4848}        
& \multicolumn{1}{c}{0.1320} 
& \multicolumn{1}{c}{0.3095} & \multicolumn{1}{c}{0.3095}
& \multicolumn{1}{c}{0.3939}& 0.1212   \\ \hline
\hline
 \multicolumn{8}{|c|}{\textbf{G5}} \\ \hline
{\textbf{0.5}} & \multicolumn{1}{c}{0.3818}  
& \multicolumn{1}{c}{0.5476}
& \multicolumn{1}{c}{0.3810}       
& \multicolumn{1}{c}{0.4902}    
& \multicolumn{1}{c}{0.2619}
& \multicolumn{1}{c}{{\color[HTML]{036400} \textbf{0.0238}}}& {{\color[HTML]{036400} \textbf{0.0238}}}          \\ \hline
{\textbf{0.6}} & \multicolumn{1}{c}{0.9048}  
& \multicolumn{1}{c}{0.5476}
& \multicolumn{1}{c}{0.2619}     
& \multicolumn{1}{c}{{\color[HTML]{036400} \textbf{0.0238}}}  
& \multicolumn{1}{c}{0.3810}
& \multicolumn{1}{c}{0.1667}  & 0.2619       \\ 
\hline
{\textbf{0.7}} & \multicolumn{1}{c}{0.3810}  
& \multicolumn{1}{c}{0.3810}
& \multicolumn{1}{c}{0.5476}      
& \multicolumn{1}{c}{0.2619}    
& \multicolumn{1}{c}{0.3810}
& \multicolumn{1}{c}{0.2619}& 0.1667        \\ \hline
{\textbf{0.8}} & \multicolumn{1}{c}{0.5476}  
& \multicolumn{1}{c}{0.5476}
& \multicolumn{1}{c}{1}      
& \multicolumn{1}{c}{0.7143}    
& \multicolumn{1}{c}{0.7143} 
& \multicolumn{1}{c}{0.7143} & 0.0952       \\ \hline
{\textbf{0.9}}  & \multicolumn{1}{c}{0.7143}
& \multicolumn{1}{c}{0.9048}        
& \multicolumn{1}{c}{0.9048}  
& \multicolumn{1}{c}{1} & \multicolumn{1}{c}{0.5476}
& \multicolumn{1}{c}{0.9048}& 0.1667  \\ \hline
\hline
 \multicolumn{8}{|c|}{\textbf{G6}}      \\ \hline
{\textbf{0.5}} & \multicolumn{1}{c}{0.4286} 
& \multicolumn{1}{c}{0.6429}
& \multicolumn{1}{c}{0.4286}        
& \multicolumn{1}{c}{0.1144}    
& \multicolumn{1}{c}{0.4286}
& \multicolumn{1}{c}{0.0714}& 0.0714      \\ \hline
{\textbf{0.6}} & \multicolumn{1}{c}{1}  
& \multicolumn{1}{c}{0.2857}
& \multicolumn{1}{c}{0.6429}  
& \multicolumn{1}{c}{0.0714}  
& \multicolumn{1}{c}{0.6429}
& \multicolumn{1}{c}{1}& 0.0714      \\ 
\hline
{\textbf{0.7}} & \multicolumn{1}{c}{0.2857} 
& \multicolumn{1}{c}{0.2857}
& \multicolumn{1}{c}{0.0714}      
& \multicolumn{1}{c}{0.2857}    
& \multicolumn{1}{c}{0.8571}
& \multicolumn{1}{c}{0.2857}& 0.2857       \\ \hline
{\textbf{0.8}} & \multicolumn{1}{c}{0.4021}  
& \multicolumn{1}{c}{0.7716}
& \multicolumn{1}{c}{0.8727}     
& \multicolumn{1}{c}{0.5483}    
& \multicolumn{1}{c}{0.5142}
& \multicolumn{1}{c}{0.5114}& 0.1489        \\ \hline
{\textbf{0.9}}  & \multicolumn{1}{c}{0.5460} 
& \multicolumn{1}{c}{0.6030}        
& \multicolumn{1}{c}{0.2057}   
& \multicolumn{1}{c}{0.5208} & \multicolumn{1}{c}{0.6983}
& \multicolumn{1}{c}{0.576}& 0.1462  \\ \hline
\noalign {\hrule height 1pt }
\end{tabular}
\end{tiny}
}
\end{table*}

Table \ref{tab:featuresTotal1} presents the separability results for groups G1–G6 using the proposed method. Spectral features are computed over spectrograms weighted by occlusion maps. The $p$-values are reported for different threshold values (Th.), specifically $\{0.5, 0.6, 0.7, 0.8, 0.9\}$. Statistically significant results ($p < 0.05$) are observed across various thresholds and features. The most discriminative features are those with Th. $= 0.7$, where all features, except $SpFx$, achieve statistical significance for groups G2--G4. 

\begin{figure*}[htb!]
   \includegraphics[width=0.5\textwidth]{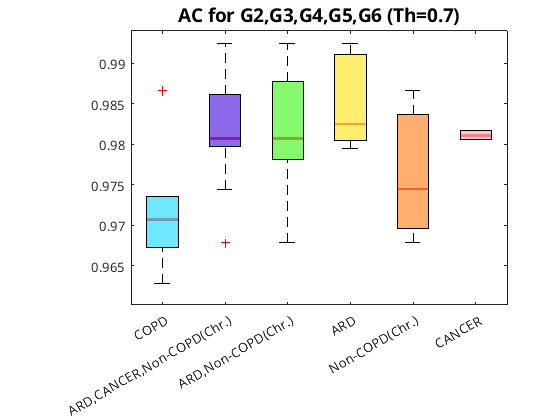}
   \includegraphics[width=0.5\textwidth]{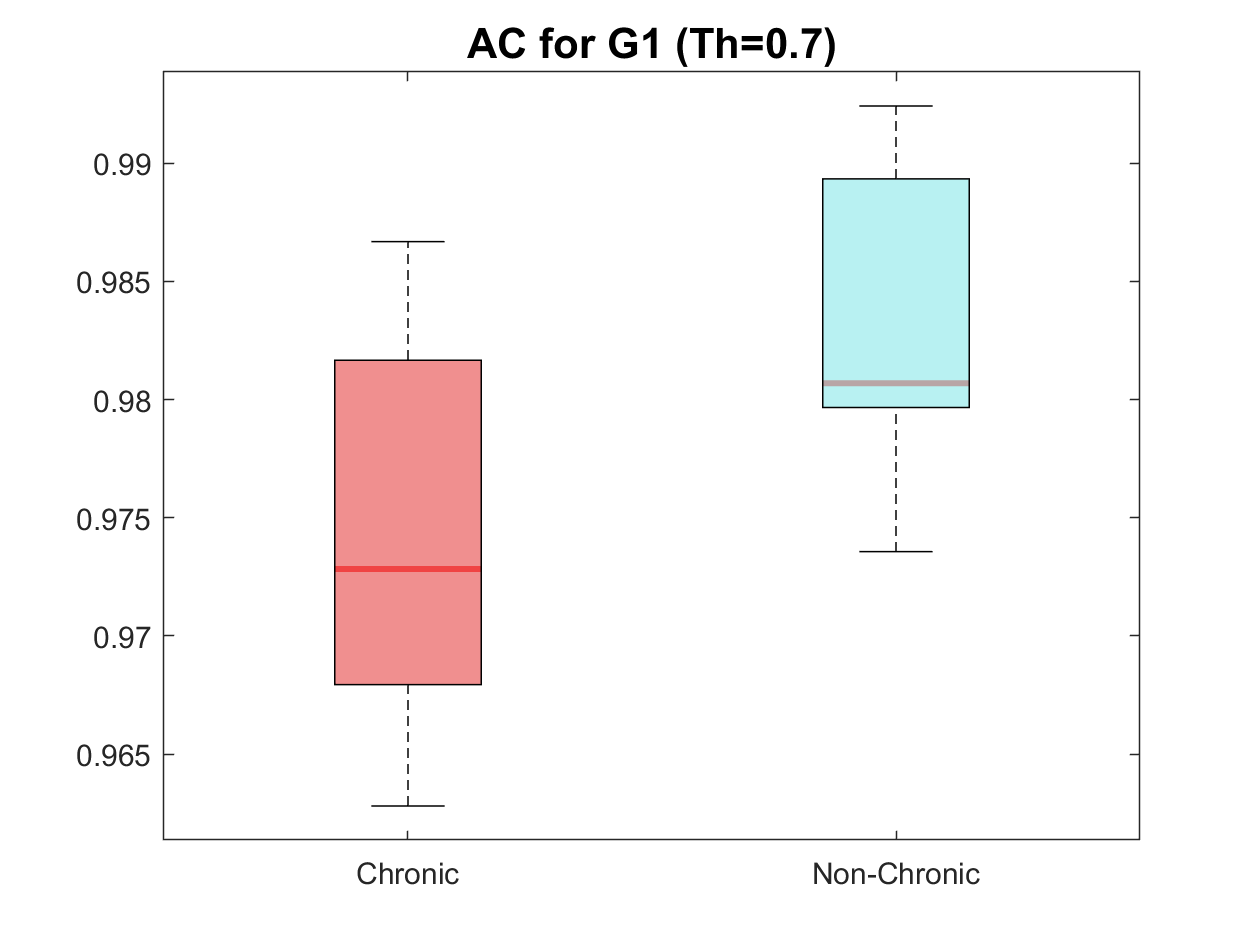}
\caption{Boxplots obtained for Relative AC Power ($AC$) at the most discriminative threshold.}
\label{fig:ACBox}
\end{figure*}

Figure \ref{fig:ACBox} presents boxplots for AC power ($AC$) at the most discriminative threshold. The boxplots on the left-hand side reveal that COPD patients show significantly lower values for this feature compared to other diseases (G2), other diseases excluding cancer (G3), and ARD/pneumonia (G4). In lung cancer patients (G6), the boxplots reveal noticeable differences, but statistical significance was not achieved due to the small sample size. Lower AC power indicates a higher DC component in the weighted power spectrum, which typically corresponds to a less variable cough pattern. However, as shown by the occlusion maps (Figure \ref{figweightedspectrograms}), most of the power in the weighted spectrogram corresponds to AC power. As a result, differences in AC power are subtle but still relevant. The boxplots in the right-hand side of the figure exhibit considerable overlap, consistent with the non-significant results found for $AC$ in group G1 (chronic vs. non-chronic patients). The same happens for the comparison between COPD and other chronic diseases (non-COPD) in group G5 (left-hand side).

\begin{figure*}[htb!]
   \includegraphics[width=0.5\textwidth]{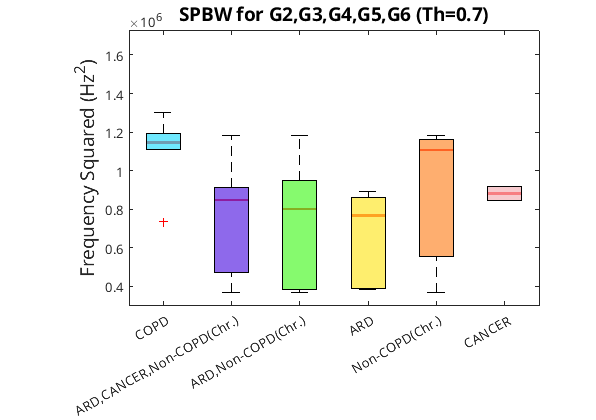}
    \includegraphics[width=0.5\textwidth]{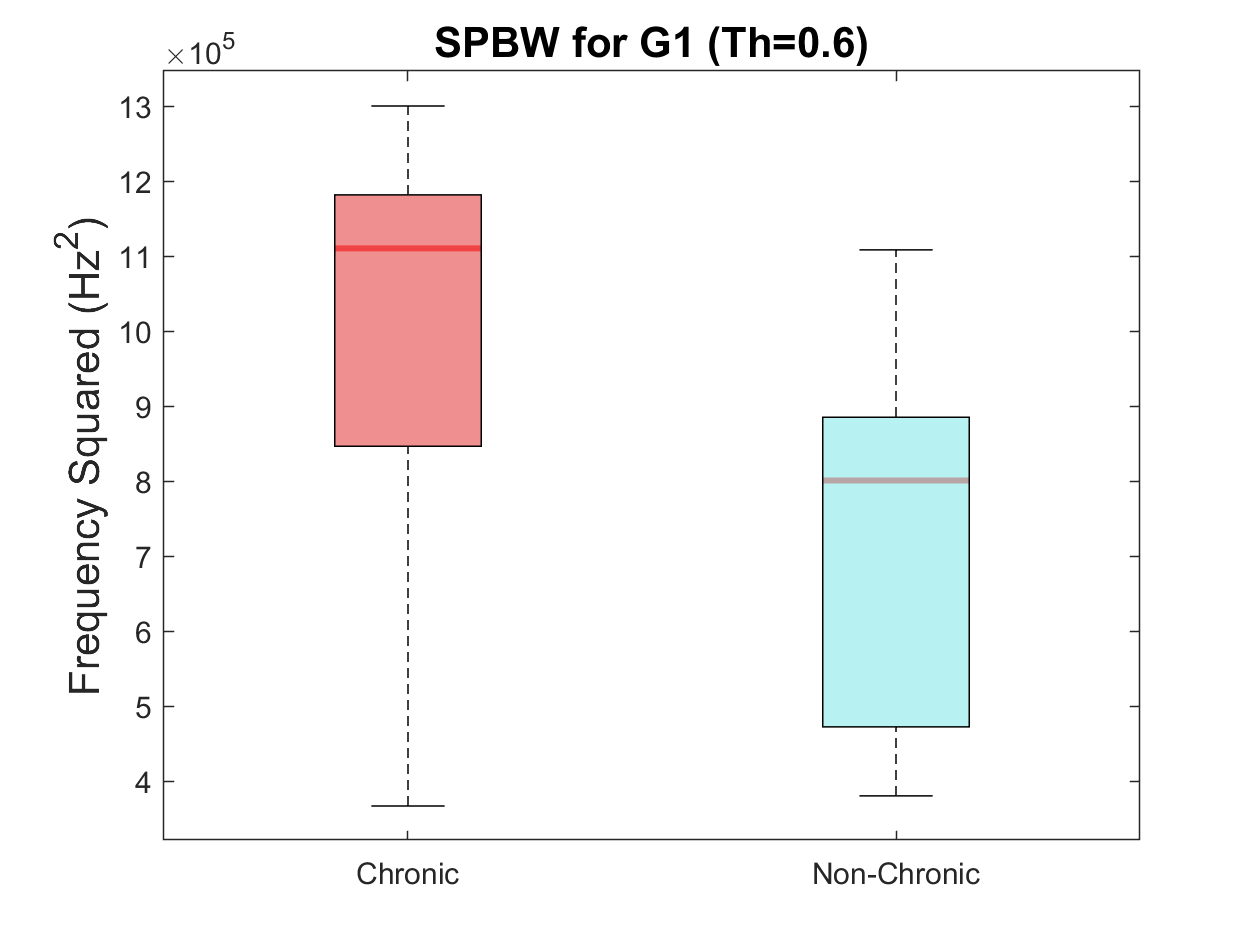}
\caption{Boxplots obtained for Spectral Bandwidth ($SpBw$) at the most discriminative threshold.}
\label{fig:SPBWBox}
\end{figure*}

Figure \ref{fig:SPBWBox} shows boxplots for Spectral Bandwidth ($SpBw$) at the most discriminative threshold. In this case, a higher value for COPD patients is observed with no overlap for groups G2--G4 (other diseases, other diseases excluding cancer, and ARD/peumonia, see left-hand side figure), which is consistent with $p$-values in table \ref{tab:featuresTotal1}. For lung cancer patients (G6), no overlap in the boxplots is observed either but again, $p>0.05$ is obtained due to the small sample size. The higher bandwidth values for COPD patients indicate a more dispersed power distribution in the weighted spectrograms, likely due to the absence of cough energy in specific frequency bands (e.g., around $1.5$ kHz), where other conditions exhibit stronger presence. This trend is visually evident in Figure \ref{figweightedspectrograms}, where the weighted spectrogram of a COPD patient differs from those of other groups. On the other hand, boxplots for G5 (figure in the left-hand side) and G1 (right hand-side) overlap significantly, as could be expected from the obtained $p$-values.

\begin{figure*}[h!]
   \includegraphics[width=0.33\textwidth]{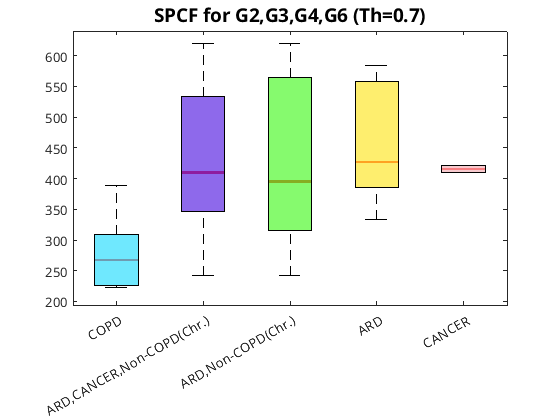}
   \includegraphics[width=0.33\textwidth]{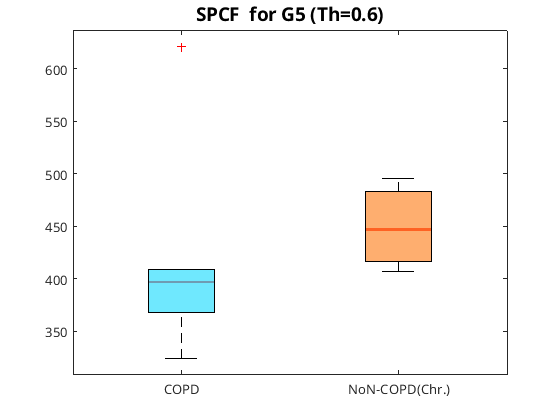}
   \includegraphics[width=0.33\textwidth]{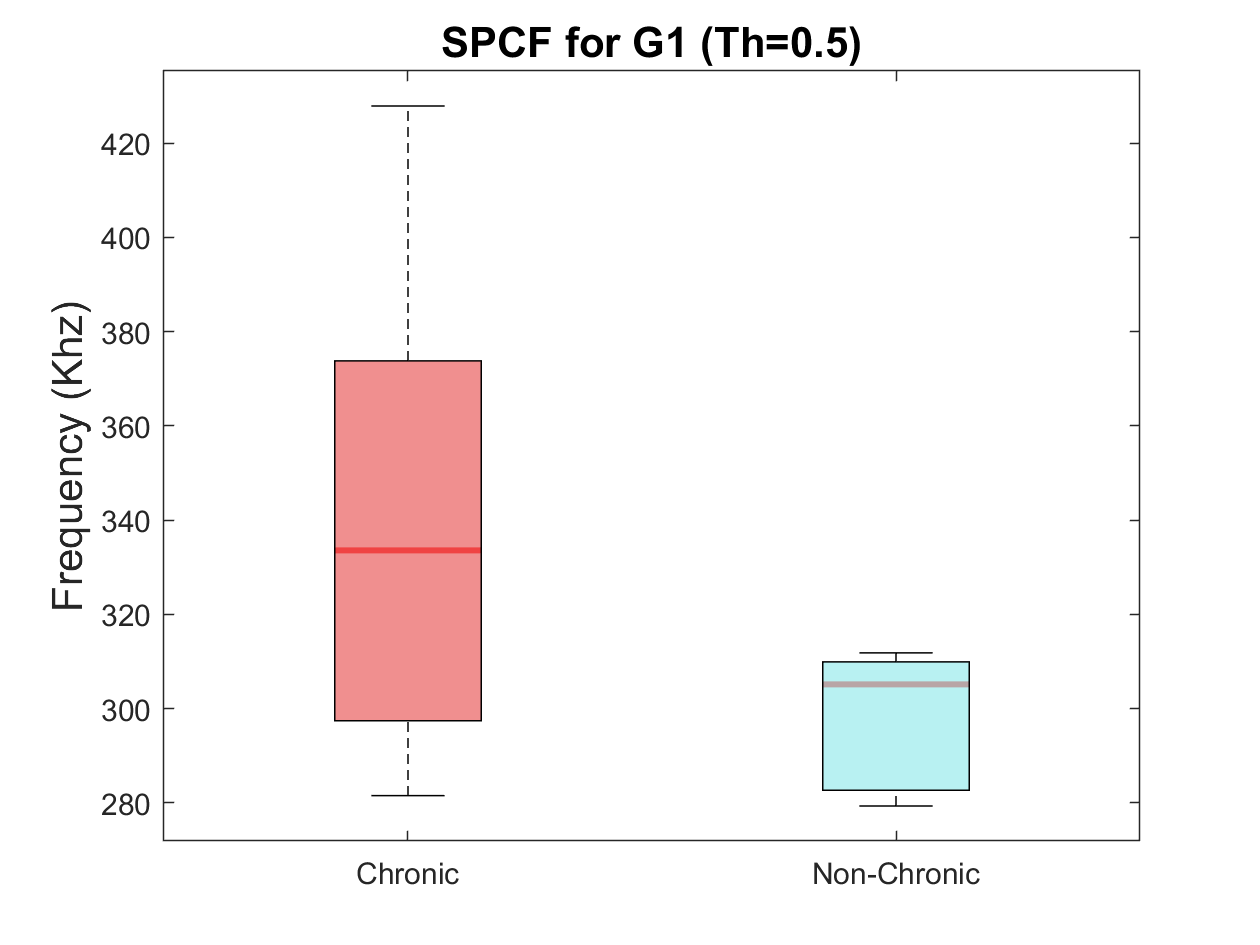}
\caption{Boxplots obtained for Spectral Crest Factor ($SpCF$) at the most discriminative threshold.}
\label{fig:SPCFBox}
\end{figure*}

Similar to $AC$, Spectral Crest Factor ($SpCF$, see Figure \ref{fig:SPCFBox}, left-hand side) exhibits significantly lower values for the COPD group, indicating a lower frequency as the "center of gravity" compared to groups G2–G4. This trend is supported by $p$-values below $0.05$, with no overlap in the boxplots. Likewise, no overlap is observed for lung cancer patients (G6); however, the $p$-value of $0.0714$ suggests that while the difference is notable, it does not yet reach statistical significance. As shown in the middle figure, no significant difference is found for G5. In contrast, chronic and non-chronic patients (G1, see right-hand side figure) exhibit a statistically significant difference ($p = 0.0330$), with non-chronic patients displaying lower $SpCF$ values at Th.$=0.5$.

\begin{figure*}[h!]
   \includegraphics[width=0.5\textwidth]{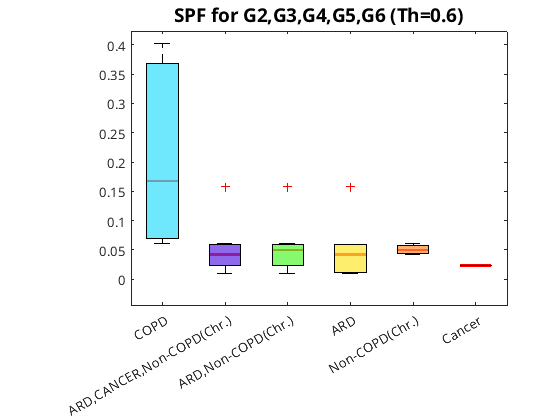}
   \includegraphics[width=0.5\textwidth]{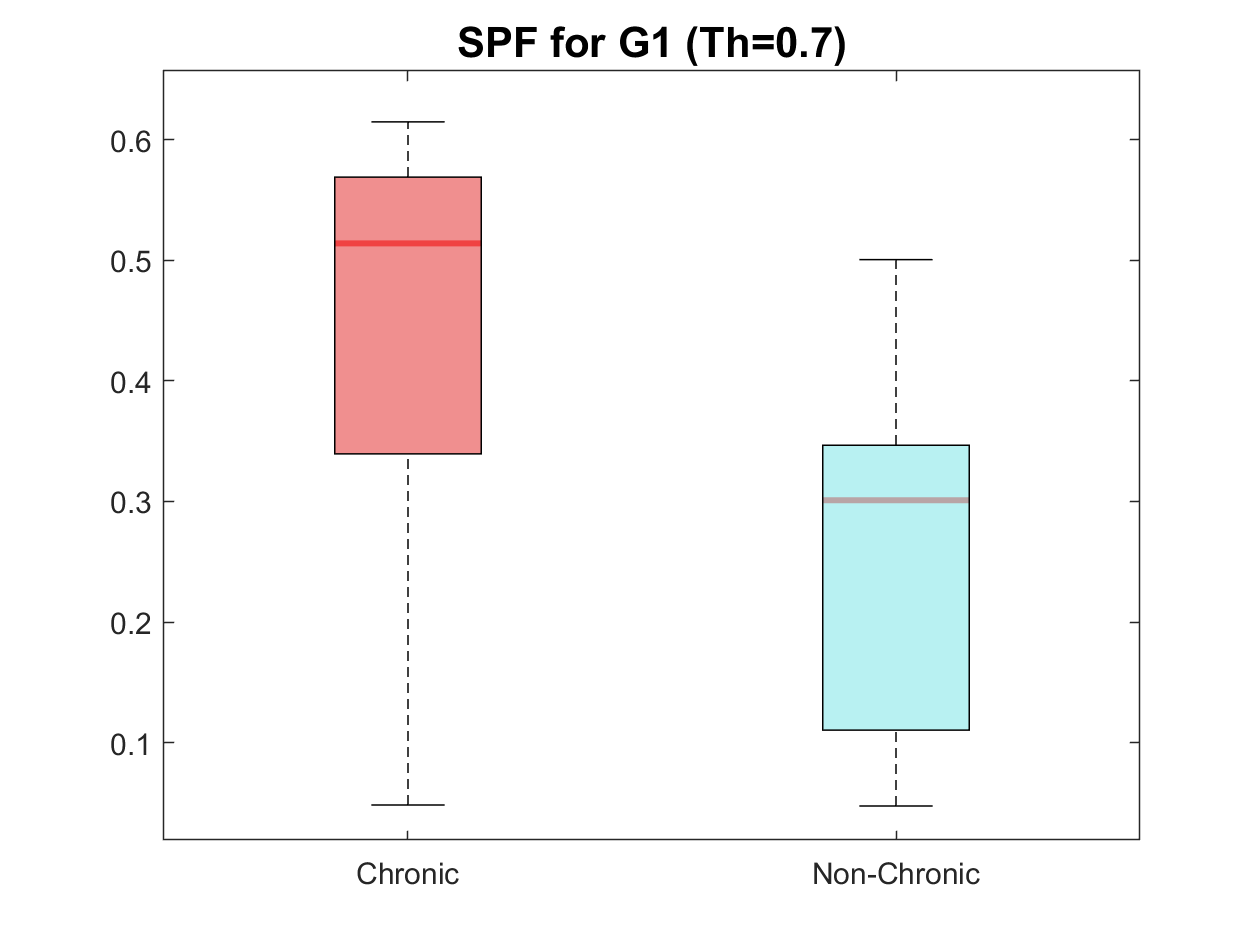}
\caption{Boxplots obtained for Spectral Flatness ($SpF$) at the most discriminative threshold.}
\label{fig:SPFBox}
\end{figure*}

Spectral Flatness ($SpF$) emerges as a highly discriminative feature, as shown in Figure \ref{fig:SPFBox}. COPD patients exhibit higher $SpF$ values, with no overlap in the boxplots compared to groups G2–G5 ($p < 0.05$) and G6 ($p = 0.0714$), as can be seen in the left-hand side of the figure. This suggests that COPD-weighted spectrograms display flatter spectra, resembling a more white noise-like distribution compared to other diseases. Regarding G1, chronic patients also show flatter spectra than non-chronic ones (see right-hand side of the figure), with $p = 0.053$, indicating a trend toward statistical significance.

\begin{figure*}[h!]
   \includegraphics[width=0.33\textwidth]{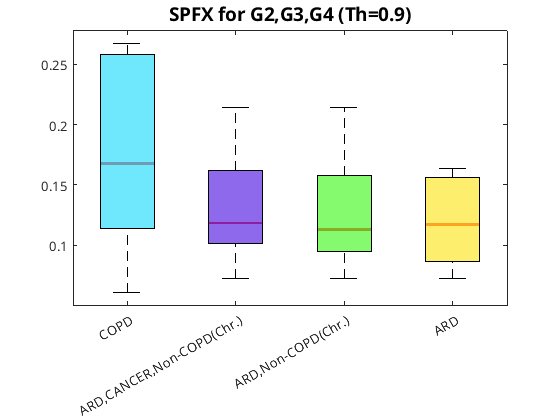}
   \includegraphics[width=0.33\textwidth]{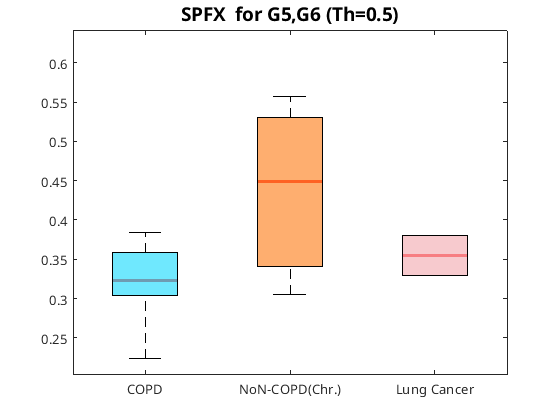}
   \includegraphics[width=0.33\textwidth]{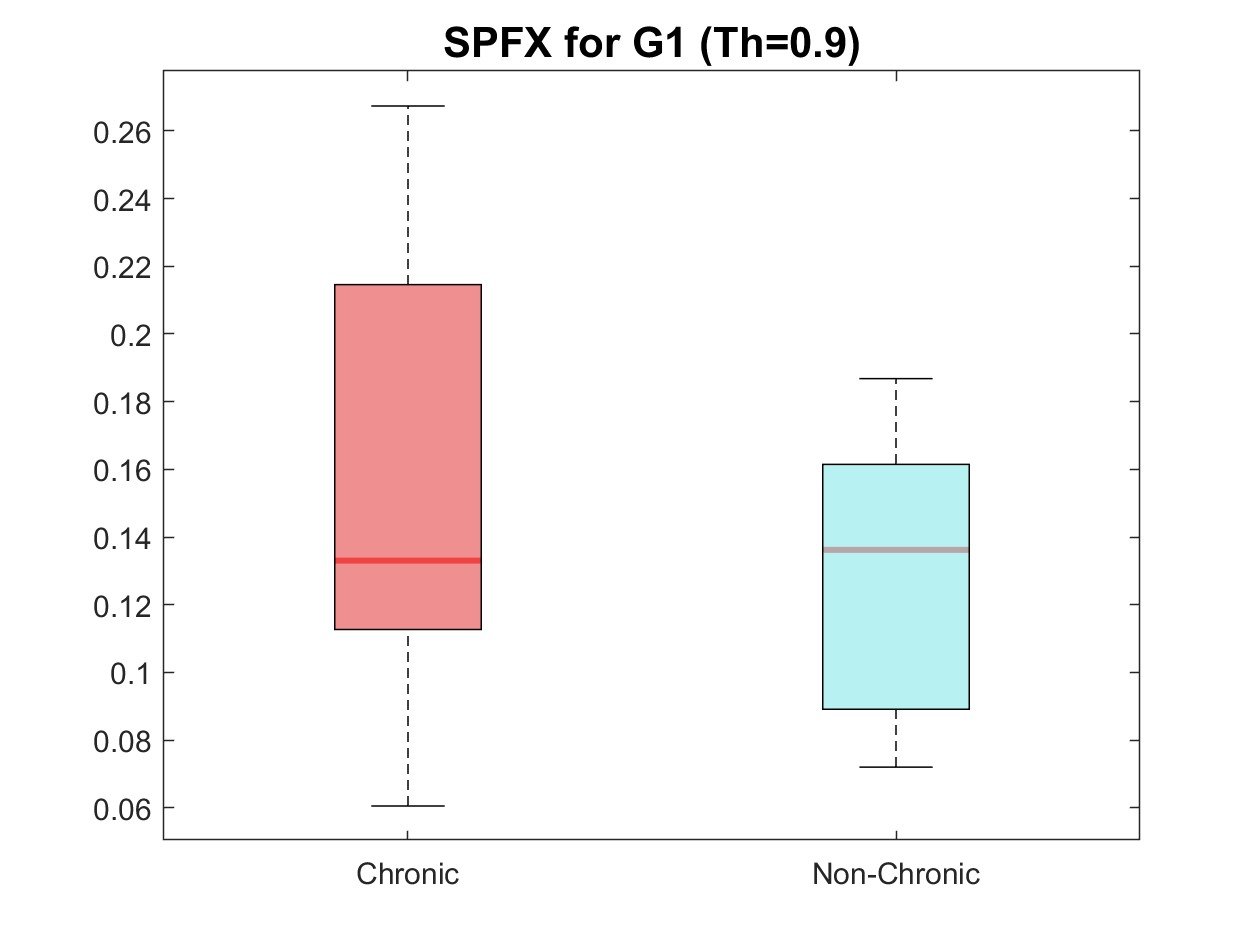}
\caption{Boxplots obtained for Spectral Flux ($SpFx$) at the most discriminative threshold.}
\label{fig:SPFXBox}
\end{figure*}

Boxplots for Spectral Flux ($SpFx$) are shown in Figure \ref{fig:SPFXBox}, where considerable overlap among groups can be observed. This aligns with the lack of statistical significance reported in Table \ref{tab:featuresTotal1}, suggesting that $SpFx$ is not  a differentiating feature among the studied conditions.

\begin{figure*}[h!]
   \includegraphics[width=0.33\textwidth]{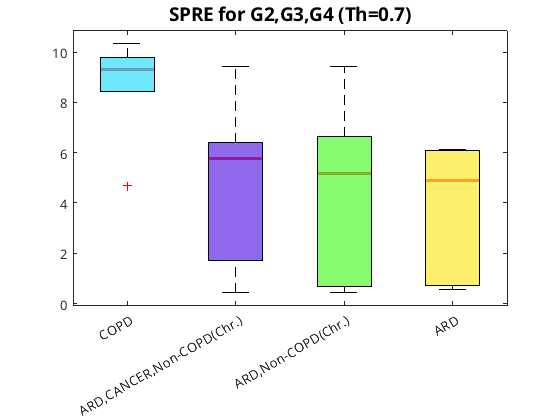}
   \includegraphics[width=0.33\textwidth]{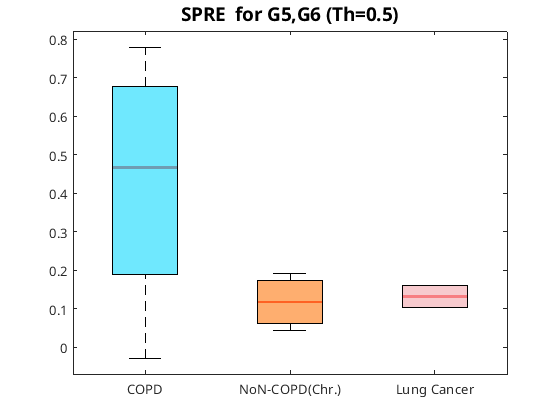}
   \includegraphics[width=0.33\textwidth]{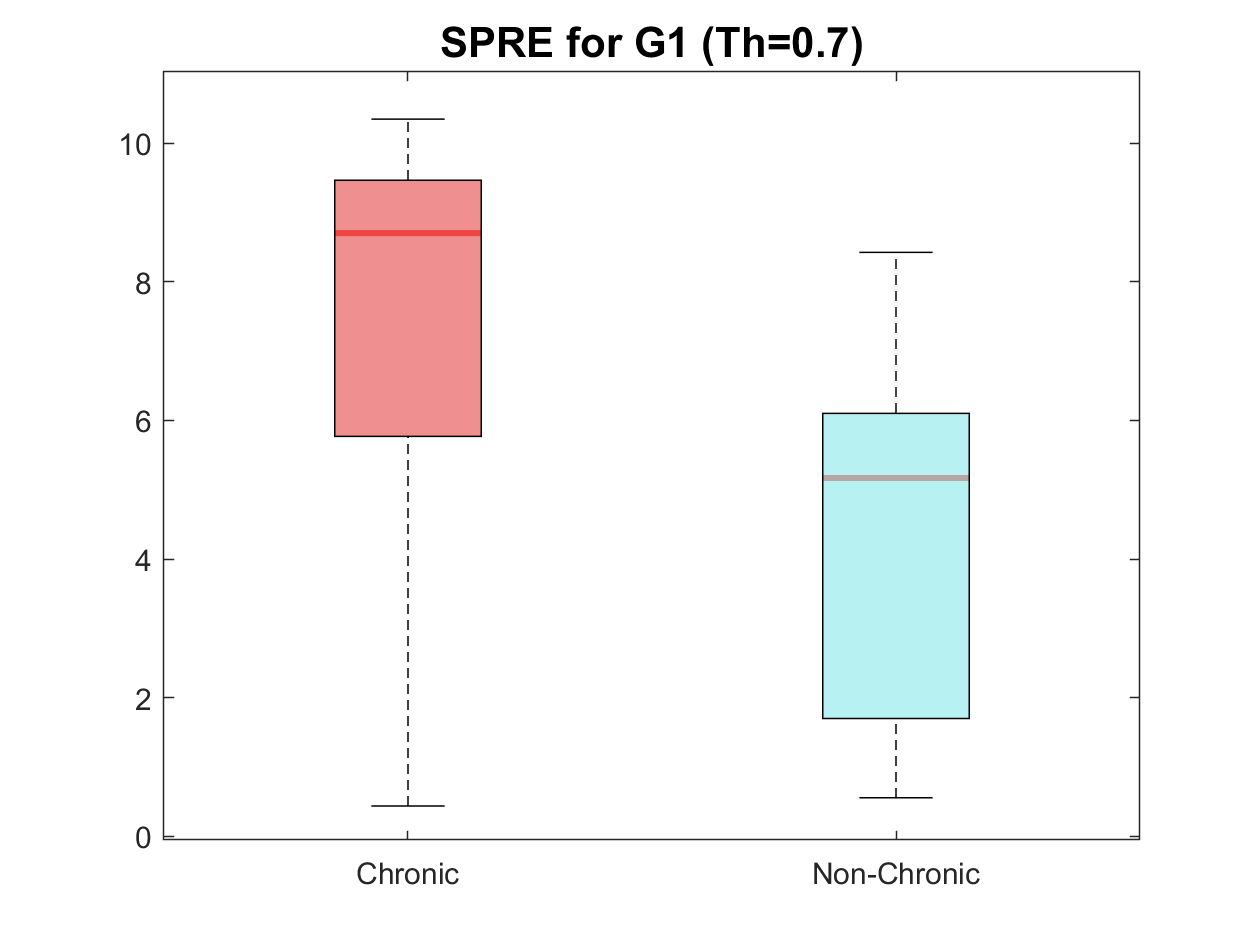}
\caption{Boxplots obtained for Spectral Renyi Entropy ($SpRE$) at the most discriminative threshold.}
\label{fig:SPREBox}
\end{figure*}

Spectral Renyi Entropy ($SpRE$) also proves to be a highly discriminative feature for groups G2–G6, as evidenced by the non-overlapping boxplots in Figure \ref{fig:SPREBox} (left and middle). COPD patients exhibit significantly higher values ($p < 0.05$) compared to other diseases, except for lung cancer ($p = 0.0714$). For G1 (right-hand side of the figure), chronic patients also show close to significant, higher $SpRE$ values than non-chronic ones ($p=0.0702$). This feature reflects a general measure of uncertainty or randomness in the spectra, suggesting that COPD and chronic patients exhibit more randomness in their frequency patterns. A higher SpRE usually means that the signal has a wide distribution of frequencies with less concentration in specific frequencies. This is often seen in signals that are broadband or noisy in nature, which also tends to increase spectral flatness. This observation is thus consistent with the results observed for $SpF$.

\begin{figure*}[h!]
   \includegraphics[width=0.33\textwidth]{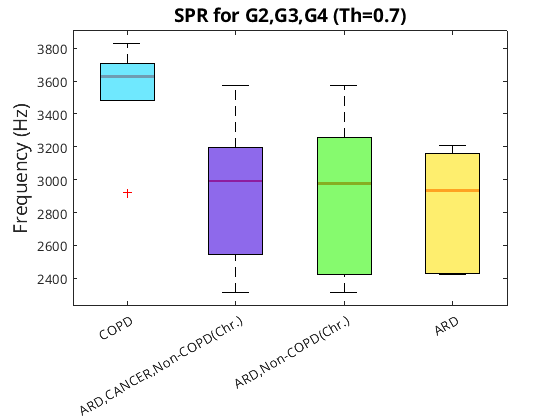}
   \includegraphics[width=0.33\textwidth]{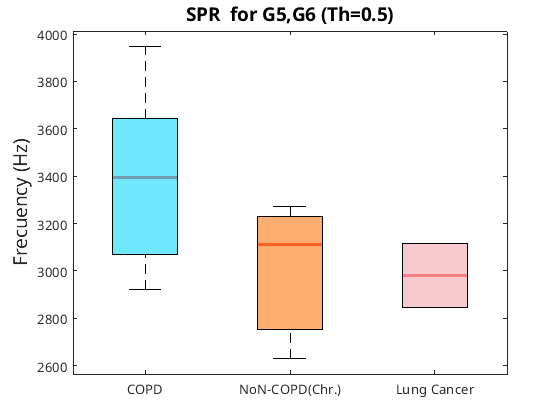}
      \includegraphics[width=0.33\textwidth]{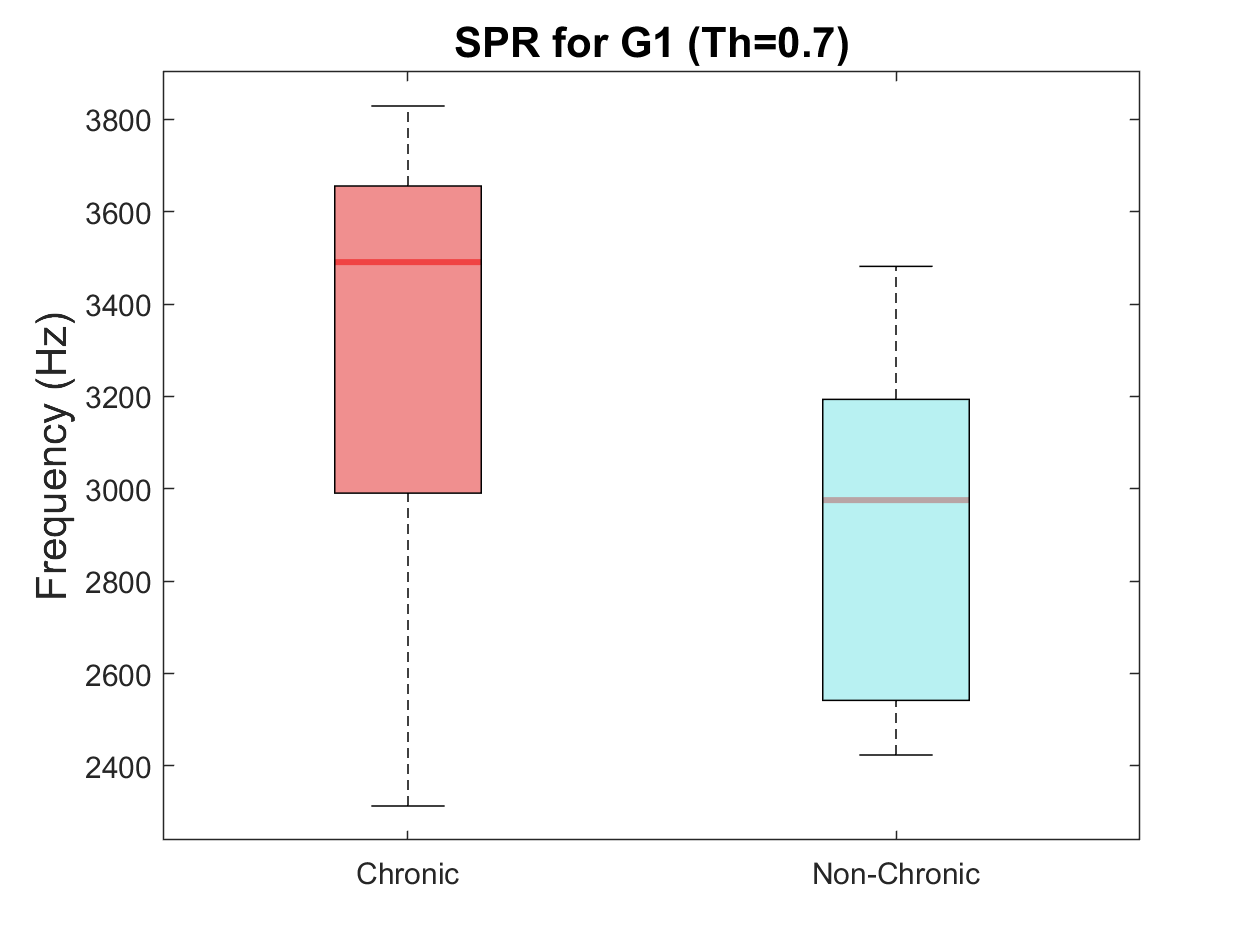}
\caption{Boxplots obtained for Spectral Roll-Off ($SpR$) at the most discriminative threshold.}
\label{fig:SPRBox}
\end{figure*}

Figure \ref{fig:SPRBox} presents boxplots for Spectral Roll-Off ($SpR$) at the most discriminative thresholds. This feature represents the frequency below which 85\% of the total power is accumulated. As shown in the left and middle subfigures, COPD patients exhibit significantly higher $SpR$ values than those in groups G2–G5 ($p < 0.05$) and G6 ($p = 0.0714$). Similarly, for G1 (right), chronic patients tend to have higher $SpR$ values than non-chronic ones, with results nearing statistical significance. These findings suggest greater variability in COPD and chronic cough patterns, as their weighted spectra extend to higher frequencies.

\begin{table*}[!htb]
\caption{Separability results obtained for groups G1--G6 using spectral features computed over original, non-weighted spectrograms. Statistical significance is considered for $p<0.05$.}
\label{tab:featuresTNoWeight}
\resizebox{\textwidth}{!} {

\begin{tiny}
\begin{tabular}{|c|ccccccc|}
\noalign{\hrule height 1pt }
\cline{2-8}
\multicolumn{1}{|l|}{}                                        & \multicolumn{1}{c|}{AC}                             & \multicolumn{1}{c|}{\textbf{$SpBW$}}                       & \multicolumn{1}{c|}{\textbf{$SpCF$}}                                           & \multicolumn{1}{c|}{\textbf{$SpF$}}                       & 
\multicolumn{1}{c|}{\textbf{$SpFx$}}                       & 
\multicolumn{1}{c|}{\textbf{$SpRE$}}                       & \textbf{$SpR$}                       \\ \cline{2-8}
\hline \hline 

{\textbf{G1}} & \multicolumn{1}{c}{0.6009}  
& \multicolumn{1}{c}{0.8868}
& \multicolumn{1}{c}{0.3638}     
& \multicolumn{1}{c}{0.5362}    
& \multicolumn{1}{c}{0.7396}
& \multicolumn{1}{c}{0.6691}&1  \\ \hline
{\textbf{G2}} & \multicolumn{1}{c}{0.1490}  
& \multicolumn{1}{c}{0.2161}
& \multicolumn{1}{c}{0.4043}     
& \multicolumn{1}{c}{0.9612}    
& \multicolumn{1}{c}{0.2561}
& \multicolumn{1}{c}{0.2561}&0.2561    \\ 
\hline
{\textbf{G3}} & \multicolumn{1}{c}{0.1135}  
& \multicolumn{1}{c}{0.3884}
& \multicolumn{1}{c}{0.3277}      
& \multicolumn{1}{c}{0.9546}    
& \multicolumn{1}{c}{0.3884}
& \multicolumn{1}{c}{0.3884}&  0.3884  \\ \hline
{\textbf{G4}} & \multicolumn{1}{c}{0.2403}  
& \multicolumn{1}{c}{0.5887}
& \multicolumn{1}{c}{0.1797}        
& \multicolumn{1}{c}{0.9372}    
& \multicolumn{1}{c}{0.4848}
& \multicolumn{1}{c}{0.9144}&0.4848\\ \hline
{\textbf{G5}} & \multicolumn{1}{c}{0.1667}  
& \multicolumn{1}{c}{0.3810}
& \multicolumn{1}{c}{1}      
& \multicolumn{1}{c}{0.7143}    
& \multicolumn{1}{c}{0.5476}
& \multicolumn{1}{c}{0.1454}& 0.5476   \\ \hline
{\textbf{G6}}  & \multicolumn{1}{c}{0.8571} 
& \multicolumn{1}{c}{0.1429}        
& \multicolumn{1}{c}{1}    
& \multicolumn{1}{c}{1}  
& \multicolumn{1}{c}{0.2857} & \multicolumn{1}{c}{0.3129}
& 0.2857 \\  \hline

\noalign {\hrule height 1pt }
\end{tabular}
\end{tiny}
}
\end{table*}

\begin{table*}[!htb]
\caption{Comparison of minimum $p$-values obtained for groups G1--G6 using spectral features computed over XAI-driven weighted spectrograms and GMM model parameters proposed in \cite{amado2024}. Statistical significance is considered for $p<0.05$. Best result for each comparison group is highlighted using green, boldfaced font.}
\label{tab:tablaComparativa}
\resizebox{\textwidth}{!} {

\begin{tiny}
\begin{tabular}{|c|cccccccc|}
\noalign{\hrule height 1pt }

\multicolumn{1}{|l|}{}                                        & \multicolumn{2}{c|}{\textbf{GMM Model} ~\cite{amado2024}}                                             & \multicolumn{6}{c|}{\textbf{XAI-driven spectral analysis (this work)}}                                              

\\
\noalign{\hrule height 0.5pt }
\multicolumn{1}{|l|}{}                                        & \multicolumn{1}{c}{\textbf{$\eta_x$ }  }                           & \multicolumn{1}{c|}{\textbf{$\sigma_x$ }}                       & \multicolumn{1}{c}{AC}                       & \multicolumn{1}{c}{\textbf{$SpBW$}}                       & \multicolumn{1}{c}{\textbf{$SpCF$}}                       & 
\multicolumn{1}{c}{\textbf{$SpF$}}                       & 
\multicolumn{1}{c}{\textbf{$SpRE$}}                       & \textbf{$SpR$}                           \\ \cline{2-9}
\hline

{\textbf{G1}} & \multicolumn{1}{c}{0.1088}  
& \multicolumn{1}{c|}{0.4747}
& \multicolumn{1}{c}{0.1088}
& \multicolumn{1}{c}{0.0702}        
& \multicolumn{1}{c}{\color[HTML]{036400} \textbf{0.0330}}       
& \multicolumn{1}{c}{0.0553}
& \multicolumn{1}{c}{0.0702}&0.0878 \\ \hline

{\textbf{G2}} & \multicolumn{1}{c}{0.0019}  
& \multicolumn{1}{c|}{0.0145}
& \multicolumn{1}{c}{0.0273}
& \multicolumn{1}{c}{0.0202}        
& \multicolumn{1}{c}{0.0103}    
& \multicolumn{1}{c}{\color[HTML]{036400} \textbf{0.0011}}
& \multicolumn{1}{c}{0.0145}&0.0103 \\ \hline
{\textbf{G3}} & \multicolumn{1}{c}{0.0048}  
& \multicolumn{1}{c|}{0.0360}
& \multicolumn{1}{c}{0.0360}
& \multicolumn{1}{c}{0.0256}        
& \multicolumn{1}{c}{0.0256}
& \multicolumn{1}{c}{\color[HTML]{036400} \textbf{0.0028}}   
& \multicolumn{1}{c}{0.0176}& {0.0120}    \\ 
\hline
{\textbf{G4}} & \multicolumn{1}{c}{0.0260}  
& \multicolumn{1}{c|}{0.0411}
& \multicolumn{1}{c}{0.0260}
& \multicolumn{1}{c}{0.0152}      
& \multicolumn{1}{c}{\color[HTML]{036400} \textbf{0.0087}}       
& \multicolumn{1}{c}{0.0152}
& \multicolumn{1}{c}{0.0152}& 0.0152  \\ \hline

{\textbf{G5}} & \multicolumn{1}{c}{\color[HTML]{036400} \textbf{0.0238}}
& \multicolumn{1}{c|}{0.3686}
& \multicolumn{1}{c}{0.3810}
& \multicolumn{1}{c}{0.3818}      
& \multicolumn{1}{c}{0.2619}    
& \multicolumn{1}{c}{\color[HTML]{036400} \textbf{0.0238}}
& \multicolumn{1}{c}{\color[HTML]{036400} \textbf{0.0238}}& {\color[HTML]{036400} \textbf{0.0238}} \\ \hline

{\textbf{G6}} & \multicolumn{1}{c}{0.0199}  
& \multicolumn{1}{c|}{\color[HTML]{036400} \textbf{0.0139}}   
& \multicolumn{1}{c}{0.2857}
& \multicolumn{1}{c}{0.2857}     
& \multicolumn{1}{c}{0.0714}
& \multicolumn{1}{c}{0.0714}
& \multicolumn{1}{c}{0.0714}& 0.0714         \\ \hline

\noalign {\hrule height 1pt }
\end{tabular}
\end{tiny}
}
\end{table*}

For the sake of comparison, table \ref{tab:featuresTNoWeight} reports the obtained $p$-values for spectral features computed over the original, non-weighted spectrogram. No feature achieved statistical significance, highlighting the effectiveness of using XAI-driven weighted spectrograms.

Finally, table \ref{tab:tablaComparativa} compares the group separability results for G1–G6 obtained in this study with those reported in \cite{amado2024}. The table shows the minimum $p$-values for each spectral feature alongside those from \cite{amado2024}, where the most discriminative features were the temporal mean $\eta_x$ and standard deviation $\sigma_x$ of the non-dominant Gaussian in the GMM model. The spectral features used in this work outperform the temporal features from \cite{amado2024} in all cases except for G6, where no significant differences were found. However, as discussed, group separability remains evident, as there is no overlap in the boxplots shown in Figures \ref{fig:ACBox}–\ref{fig:SPRBox}.


\section{Conclusion}
\label{conclusion}
This paper proposes a novel XAI-based methodology to improve the understanding of cough sound analysis in respiratory disease management. By using occlusion maps to highlight relevant regions in spectrograms processed by a CNN-based cough detector, the proposed approach enables a more interpretable feature extraction process. The subsequent spectral analysis demonstrates significant differences between disease groups when features are derived from spectrograms weighted by the occlusion maps, whereas such differences remain undetectable in raw spectrograms alone. Specifically, cough patterns appear as more variable in the spectral regions of interest in patients suffering from COPD compared to other disease groups. These findings underscore the potential of XAI-driven techniques to enhance the diagnostic capabilities of cough sound analysis.

\section*{CRediT authorship contribution statement}

\textbf{P.A.C.}: Methodology, Data Curation, Formal Analysis, Writing -- Original Draft. \textbf{L.S.R.}: Investigation, Methodology, Writing -- Review \& Editing.  \textbf{M.A.G.}: Investigation, Resources, Validation. \textbf{J.G.L.}: Investigation, Resources, Validation, Funding Acquisition. \textbf{C.A.L.}: Methodology, Supervision, Writing -- Review \& Editing. \textbf{P.C.H.}: Conceptualization, Writing -- Review \& Editing, Supervision, Funding Acquisition.

\section*{Ethics statement}
The study was carried out in accordance with the Declaration of Helsinki and was approved by the Área de Salud de Palencia Research Ethics Committee meetings 17/05/2018 and 09/08/2023. Subjects provided their informed consent before the recordings.

\section*{Conflict of interest statement}
The authors declare no conflict of interest.
\section*{Funding}
This work was supported by projects TED2021-131536B-I00, PID2022-142166NA-I0, and CPP2021-008880, funded by the Spanish MCIN/AEI/10.13039/501100011033, with TED2021-131536B-I00 and CPP2021-008880 co-funded by the EU NextGenerationEU/PRTR. The work was also partially funded by GRS 2837/C/2023 funded by Gerencia Regional de Salud, Junta de Castilla y León, Spain, EU Horizon 2020 Research and Innovation Programme under the Marie Sklodowska-Curie grant agreement No. 101008297. This article reflects only the authors’ view. The European Union Commission is not responsible for any use that may be made of the information it contains.







\bibliographystyle{elsarticle-num} 
\bibliography{biblio_abbreviated}

@article {Belli20,
	author = {S. Belli \textit{et al.}},
	title = {Low physical functioning and impaired performance of activities of daily life in {COVID-19} patients who survived hospitalisation},
	volume = {56},
	number = {4},
	elocation-id = {2002096},
	year = {2020},
	journal = {Eur. Respir. J.}
}

@book{who17,
  author       = {{World Health Organisation}},
   title       = {The Global Impact of Respiratory Disease},
  publisher = {European Respiratory Society},
  year        = {2017}
}

@misc{who21,
  author       = {{World Health Organisation}},
  title        = {{WHO Coronavirus ({COVID-19}) Dashboard}},

  year         = {2021}
}

@techreport{CE18,
	Author = {{European Commission}},
	Institution = {{European Union}},
	Title = {Market study on telemedicine},
	Year = {2018}}

@techreport{Audit11,
	Author = {{Audit Scotland}},
	Institution = {{Audit Scotland}},
	Title = {{A review of Telehealth in Scotland}},
	Year = {2011}}

@article{Pinnock13,
	Author = {{H. Pinnock \textit{et al.}}},
	Journal = {BMJ},
	Publisher = {BMJ Publishing Group Ltd},
	Title = {Effectiveness of telemonitoring integrated into existing clinical services on hospital admission for exacerbation of chronic obstructive pulmonary disease: researcher blind, multicentre, randomised controlled trial},
		Volume = {347},
	Year = {2013}
}

@inproceedings{tokuda94_icslp,
  title     = {Mel-generalized cepstral analysis - a unified approach to speech spectral estimation},
  author    = {Keiichi Tokuda and Takao Kobayashi and Takashi Masuko and Satoshi Imai},
  year      = {1994},
  booktitle = {3rd International Conference on Spoken Language Processing (ICSLP 1994)},
  pages     = {1043--1046},
  issn      = {2958-1796},
}

@article{swarnkar2013automatic,
  title={Automatic identification of wet and dry cough in pediatric patients with respiratory diseases},
  author={Swarnkar, Vinayak and Abeyratne, Udantha R and Chang, Anne B and Amrulloh, Yusuf A and Setyati, Amalia and Triasih, Rina},
  journal={Ann. Biomed. Eng.},
  volume={41},
  pages={1016--1028},
  year={2013},
  publisher={Springer}
}

@inproceedings{swarnkar2013neural,
  title={Neural network based algorithm for automatic identification of cough sounds},
  author={Swarnkar, Vinayak and Abeyratne, Udantha R and Amrulloh, Yusuf and Hukins, Craig and Triasih, Rina and Setyati, Amalia},
  booktitle={2013 35th Annual International Conference of the IEEE Engineering in Medicine and Biology Society (EMBC)},
  pages={1764--1767},
  year={2013},
  organization={IEEE}
}

@article{drugman2013objective,
  title={Objective study of sensor relevance for automatic cough detection},
  author={Drugman, Thomas and Urbain, Jerome and Bauwens, Nathalie and Chessini, Ricardo and Valderrama, Carlos and Lebecque, Patrick and Dutoit, Thierry},
  journal={IEEE J. Biomed. Health Inform.},
  volume={17},
  number={3},
  pages={699--707},
  year={2013},
  publisher={IEEE}
}

@article{sterling2014automated,
  title={Automated cough assessment on a mobile platform},
  author={Sterling, Mark and Rhee, Hyekyun and Bocko, Mark},
  journal={J. Med. Eng.},
  volume={2014},
  number={1},
  pages={951621},
  year={2014},
  publisher={Wiley Online Library}
}

@article{drugman2014using,
  title={Using mutual information in supervised temporal event detection: Application to cough detection},
  author={Drugman, Thomas},
  journal={Biomed. Signal Process. Control},
  volume={10},
  pages={50--57},
  year={2014},
  publisher={Elsevier}
}

@article{amrulloh2015automatic,
  title={Automatic cough segmentation from non-contact sound recordings in pediatric wards},
  author={Amrulloh, Yusuf A and Abeyratne, Udantha R and Swarnkar, Vinayak and Triasih, Rina and Setyati, Amalia},
  journal={Biomed. Signal Process. Control},
  volume={21},
  pages={126--136},
  year={2015},
  publisher={Elsevier}
}

@inproceedings{casaseca2015effect,
  title={Effect of downsampling and compressive sensing on audio-based continuous cough monitoring},
  author={Casaseca-de-la-Higuera, Pablo and Lesso, Paul and McKinstry, Brian and Pinnock, Hilary and Rabinovich, Roberto and McCloughan, Lucy and Monge-{\'A}lvarez, Jes{\'u}s},
  booktitle={2015 37th Annual International Conference of the IEEE Engineering in Medicine and Biology Society (EMBC)},
  pages={6231--6235},
  year={2015},
  organization={IEEE}
}

@INPROCEEDINGS{monge2016effect,
  author={Monge-Alvarez, Jesús and Hoyos-Barceló, Carlos and Lesso, Paul and Escudero, Javier and Dahal, Keshav and Casaseca-de-la-Higuera, Pablo},
  booktitle={2016 38th Annual International Conference of the IEEE Engineering in Medicine and Biology Society (EMBC)},
  title={Effect of importance sampling on robust segmentation of audio-cough events in noisy environments},
  year={2016},
  volume={},
  number={},
  pages={3740-3744},
  }

@article{hoyos2018efficient,
  title={Efficient computation of image moments for robust cough detection using smartphones},
  author={Hoyos-Barcel{\'o}, Carlos and Monge-{\'A}lvarez, Jes{\'u}s and Pervez, Zeeshan and San-Jos{\'e}-Revuelta, Luis M and Casaseca-de-la-Higuera, Pablo},
  journal={Comput. Biol. Med.},
  volume={100},
  pages={176--185},
  year={2018},
  publisher={Elsevier}
}

@article{hoyos2017efficient,
  title={Efficient k-NN implementation for real-time detection of cough events in smartphones},
  author={Hoyos-Barcel{\'o}, Carlos and Monge-{\'A}lvarez, Jes{\'u}s and Shakir, Muhammad Zeeshan and Alcaraz-Calero, Jose-Mar{\'\i}a and Casaseca-de-La-Higuera, Pablo},
  journal={IEEE J. Biomed. Health Inform.},
  volume={22},
  number={5},
  pages={1662--1671},
  year={2017},
  publisher={IEEE}
}

@article{monge2018audio,
  title={Audio-cough event detection based on moment theory},
  author={Monge-Alvarez, Jesus and Hoyos-Barcel{\'o}, Carlos and Dahal, Keshav and Casaseca-de-la-Higuera, Pablo},
  journal={Appl. Acoust.},
  volume={135},
  pages={124--135},
  year={2018},
  publisher={Elsevier}
}

@article{KLCO201836,
title = {Novel computer algorithm for cough monitoring based on octonions},
journal = {Respir. Physiol. Neurobiol.},
volume = {257},
pages = {36-41},
year = {2018},
note = {Cough and Airway Defense - from neurophysiology to therapy},
issn = {1569-9048},
author = {Peter Klco and Marian Kollarik and Milos Tatar}
}

@article{monge2018robust,
  title={Robust detection of audio-cough events using local hu moments},
  author={Monge-{\'A}lvarez, Jes{\'u}s and Hoyos-Barcel{\'o}, Carlos and Lesso, Paul and Casaseca-De-La-Higuera, Pablo},
  journal={IEEE J. Biomed. Health Inform.},
  volume={23},
  number={1},
  pages={184--196},
  year={2018},
  publisher={IEEE}
}

@article{Gaj06,
  author={B. Gajic and K. K. Paliwal},
  title={Robust speech recognition in noisy environments based on subband spectral centroid histograms},
  journal={IEEE Audio, Speech, Language Process.},
  volume={14},
  number={2},
  pages={600--608},
  year={2006}
}

@article{Mammone96,
  author={R. Mammone and others},
  title={Robust speaker recognition: A feature-based approach},
  journal={IEEE Signal Process. Mag.},
  volume={13},
  number={5},
  pages={58--71},
  month={Sep},
  year={1996}
}

@article{Vizel2010,
  author={Vizel, Eldad and Yigla, Mordechai and Goryachev, Yulia and Dekel, Eyal and Felis, Vered and Levi, Hanna and Kroin, Isaac and Godfrey, Simon and Gavriely, Noam},
  title={Validation of an ambulatory cough detection and counting application using voluntary cough under different conditions},
  journal={Cough},
  volume={6},
  number={1},
  pages={3},
  year={2010},
  }

@article{Birring2008,
  author={Birring, S. S. and Fleming, T. and Matos, S. and Raj, A. A. and Evans, D. H. and Pavord, I. D.},
  title={The Leicester Cough Monitor: preliminary validation of an automated cough detection system in chronic cough},
  journal={Eur. Respir. J.},
  volume={31},
  number={5},
  pages={1013--1018},
  year={2008},
  month={May},
  eprinttype={pubmed}
}

@inproceedings{Liu2013,
  author    = {Jia-Ming Liu and Mingyu You and Guo-Zheng Li and Zheng Wang and Xianghuai Xu and Zhongmin Qiu and Wenjia Xie and Chao An and Sili Chen},
  title     = {Cough signal recognition with Gammatone Cepstral Coefficients},
  booktitle = {Proceedings of the 2013 IEEE China Summit and International Conference on Signal and Information Processing},
  pages     = {160--164},
  year      = {2013},

}

@ARTICLE{Monge19,
  author={Monge{-}{\'A}lvarez, Jes{\'u}s and Hoyos-Barcel{\'o}, Carlos and San-Jos{\'e}-Revuelta, Luis Miguel and Casaseca-de-la-Higuera, Pablo},
  journal={IEEE Trans. Biomed. Eng.},
  title={A Machine Hearing System for Robust Cough Detection Based on a High-Level Representation of Band-Specific Audio Features},
  year={2019},
  volume={66},
  number={8},
  pages={2319-2330},
}

@inproceedings{amoh2015deepcough,
  title={DeepCough: A deep convolutional neural network in a wearable cough detection system},
  author={Amoh, Justice and Odame, Kofi},
  booktitle={2015 IEEE Biomedical Circuits and Systems Conference (BioCAS)},
  pages={1--4},
  year={2015},
  organization={IEEE}
}

@misc{Uni21,
  author={{University of Cambridge}},
  title={{COVID-19 Sounds App}},
  year={2021},
  note={Last access: 03/2025}
}

@misc{Kha21,
      title={Virufy: A Multi-Branch Deep Learning Network for Automated Detection of COVID-19}, 
      author={Ahmed Fakhry and Xinyi Jiang and Jaclyn Xiao and Gunvant Chaudhari and Asriel Han and Amil Khanzada},
      year={2021},
      eprint={2103.01806},
      archivePrefix={arXiv},
      primaryClass={cs.SD},
      url={https://arxiv.org/abs/2103.01806}, 
}

@article{Feng2021DeeplearningBA,
  title={Deep-learning Based Approach to Identify {COVID}-19},
  author={Ke Feng and Fengyu He and Jessica Steinmann and Ilteris Demirkiran},
  journal={SoutheastCon 2021},
  year={2021},
  pages={1-4},

}

@article{bhattacharya2023coswara,
  author    = {Bhattacharya, D. and Sharma, N.K. and Dutta, D. and et al.},
  title     = {Coswara: A respiratory sounds and symptoms dataset for remote screening of {SARS}-CoV-2 infection},
  journal   = {Sci. Data},
  volume    = {10},
  number    = {1},
  pages     = {397},
  year      = {2023},

}

@article{pahar2020covid,
 title = {{COVID}-19 cough classification using machine learning and global smartphone recordings},
journal = {Comput. Biol. Med.},
volume = {135},
pages = {104572},
year = {2021},
author = {Madhurananda Pahar and Marisa Klopper and Robin Warren and Thomas Niesler}
}

@misc{islam2025robustcovid19detectioncough,
      title={Robust COVID-19 Detection from Cough Sounds using Deep Neural Decision Tree and Forest: A Comprehensive Cross-Datasets Evaluation}, 
      author={Rofiqul Islam and Nihad Karim Chowdhury and Muhammad Ashad Kabir},
      year={2025},
      eprint={2501.01117},
      archivePrefix={arXiv},
      primaryClass={cs.SD},
      url={https://arxiv.org/abs/2501.01117}, 
}

@article{hussain2024cough2covid19,
  author    = {Hussain, S. and Ayoub, M. and Wahid, J.A. and et al.},
  title     = {{Cough2COVID-19} detection using an enhanced multi layer ensemble deep learning framework and Cough Feature Ranker},
  journal   = {Sci. Rep.},
  volume    = {14},
  number    = {1},
  pages     = {25207},
  year      = {2024},

}

@Article{Ghrabli22,
AUTHOR = {Ghrabli, Syrine and Elgendi, Mohamed and Menon, Carlo},
TITLE = {Challenges and Opportunities of Deep Learning for Cough-Based {COVID-19} Diagnosis: A Scoping Review},
JOURNAL = {Diagnostics},
VOLUME = {12},
YEAR = {2022},
NUMBER = {9},
ARTICLE-NUMBER = {2142},

PubMedID = {36140543},
ISSN = {2075-4418},

}

@inproceedings{brown2020exploring,
  author    = {Chlo{\"e} Brown and Jagmohan Chauhan and Andreas Grammenos and et al.},
  title     = {Exploring Automatic Diagnosis of {COVID}-19 from Crowdsourced Respiratory Sound Data},
  booktitle = {Proceedings of the 26th ACM SIGKDD International Conference on Knowledge Discovery \& Data Mining},
  pages     = {7},
  year      = {2020}
}

@inproceedings{vrindavanam2021machine,
  author    = {Jayavrinda Vrindavanam and Raghunandan Srinath and Hari Haran Shankar and Gaurav Nagesh},
  title     = {Machine Learning based {COVID}-19 Cough Classification Models - A Comparative Analysis},
  booktitle = {2021 5th International Conference on Computing Methodologies and Communication (ICCMC)},
  pages     = {420--426},
  year      = {2021}
}

@article{amoh2016deep,
  title={Deep neural networks for identifying cough sounds},
  author={Amoh, Justice and Odame, Kofi},
  journal={IEEE Trans. Biomed. Circuits Syst.},
  volume={10},
  number={5},
  pages={1003--1011},
  year={2016},
  publisher={IEEE}
}

@inproceedings{liu2014cough,
  title={Cough detection using deep neural networks},
  author={Liu, Jia-Ming and You, Mingyu and Wang, Zheng and Li, Guo-Zheng and Xu, Xianghuai and Qiu, Zhongmin},
  booktitle={2014 IEEE international conference on bioinformatics and biomedicine (BIBM)},
  pages={560--563},
  year={2014},
  organization={IEEE}
}

@INPROCEEDINGS{kadambi2018,
  author={Kadambi, Prad and Mohanty, Abinash and Ren, Hao and Smith, Jaclyn and McGuinnes, Kevin and Holt, Kimberly and Furtwaengler, Armin and Slepetys, Roberto and Yang, Zheng and Seo, Jae-sun and Chae, Junseok and Cao, Yu and Berisha, Visar},
  booktitle={2018 IEEE International Conference on Acoustics, Speech and Signal Processing (ICASSP)},
  title={Towards a Wearable Cough Detector Based on Neural Networks},
  year={2018},
  volume={},
  number={},
  pages={2161-2165},
  }

@article{kvapilova2020continuous,
  title={Continuous sound collection using smartphones and machine learning to measure cough},
  author={Kvapilova, Lucia and Boza, Vladimir and Dubec, Peter and Majernik, Martin and Bogar, Jan and Jamison, Jamileh and Goldsack, Jennifer C and Kimmel, Duncan J and Karlin, Daniel R},
  journal={Digital Biomark.},
  volume={3},
  number={3},
  pages={166--175},
  year={2020},
  publisher={S. Karger AG Basel, Switzerland}
}

@article{you2022automatic,
  title={Automatic cough detection from realistic audio recordings using C-BiLSTM with boundary regression},
  author={You, Mingyu and Wang, Weihao and Li, You and Liu, Jiaming and Xu, Xianghuai and Qiu, Zhongmin},
  journal={Biomed. Signal Process. Control},
  volume={72},
  pages={103304},
  year={2022},
  publisher={Elsevier}
}

@article{abeyratne2013cough,
  title={Cough sound analysis can rapidly diagnose childhood pneumonia},
  author={Abeyratne, Udantha R and Swarnkar, Vinayak and Setyati, Amalia and Triasih, Rina},
  journal={Ann. Biomed. Eng.},
  volume={41},
  pages={2448--2462},
  year={2013},
  publisher={Springer}
}

@article{kosasih2014wavelet,
  title={Wavelet augmented cough analysis for rapid childhood pneumonia diagnosis},
  author={Kosasih, Keegan and Abeyratne, Udantha R and Swarnkar, Vinayak and Triasih, Rina},
  journal={IEEE Trans. Biomed. Eng.},
  volume={62},
  number={4},
  pages={1185--1194},
  year={2014},
  publisher={IEEE}
}

@inproceedings{amrulloh2015cough,
  title={Cough sound analysis for pneumonia and asthma classification in pediatric population},
  author={Amrulloh, Yusuf and Abeyratne, Udantha and Swarnkar, Vinayak and Triasih, Rina},
  booktitle={2015 6th international conference on intelligent systems, modelling and simulation},
  pages={127--131},
  year={2015},
  organization={IEEE}
}

@article{sharan2018automatic,
  title={Automatic croup diagnosis using cough sound recognition},
  author={Sharan, Roneel V and Abeyratne, Udantha R and Swarnkar, Vinayak R and Porter, Paul},
  journal={IEEE Trans. Biomed. Eng.},
  volume={66},
  number={2},
  pages={485--495},
  year={2018},
  publisher={IEEE}
}

@article{laguarta2020covid,
  title={{COVID}-19 artificial intelligence diagnosis using only cough recordings},
  author={Laguarta, Jordi and Hueto, Ferran and Subirana, Brian},
  journal={IEEE Open J. Eng. Med. Biol.},
  volume={1},
  pages={275--281},
  year={2020},
  publisher={IEEE}
}

@article{imran2020ai4covid,
  title={{AI4COVID-19}: {AI} enabled preliminary diagnosis for {COVID}-19 from cough samples via an app},
  author={Imran, Ali and Posokhova, Iryna and Qureshi, Haneya N and Masood, Usama and Riaz, Muhammad Sajid and Ali, Kamran and John, Charles N and Hussain, MD Iftikhar and Nabeel, Muhammad},
  journal={Inform. Med. Unlocked},
  volume={20},
  pages={100378},
  year={2020},
  publisher={Elsevier}
}

@article{tena2022automated,
  title={Automated detection of {COVID-19} cough},
  author={Tena, Alberto and Claria, Francesc and Solsona, Francesc},
  journal={Biomed. Signal Process. Control},
  volume={71},
  pages={103175},
  year={2022},
  publisher={Elsevier}
}

@article{gong2021ast,
  title={{AST}: Audio spectrogram transformer},
  author={Gong, Yuan and Chung, Yu-An and Glass, James},
  journal={arXiv preprint arXiv:2104.01778},
  year={2021}
}

@inproceedings{habashy2022cough,
  title={Cough classification using audio spectrogram transformer},
  author={Habashy, Karim and Vald{\'e}s, Julio and Cohen-McFarlane, Madison and Xi, Pengcheng and Wallace, Bruce and Goubran, Rafik and Knoefel, Frank},
  booktitle={2022 IEEE Sensors Applications Symposium (SAS)},
  pages={1--6},
  year={2022},
  organization={IEEE}
}

@article{baur2024hear,
  title={{HeAR}--Health Acoustic Representations},
  author={Baur, Sebastien and Nabulsi, Zaid and Weng, Wei-Hung and Garrison, Jake and Blankemeier, Louis and Fishman, Sam and Chen, Christina and Kakarmath, Sujay and Maimbolwa, Minyoi and Sanjase, Nsala and others},
  journal={arXiv preprint arXiv:2403.02522},
  year={2024}
}

@inproceedings{laska2024zero,
  title={Zero-Shot Multi-Task Cough Sound Analysis with Speech Foundation Model Embeddings},
  author={Laska, Brady and Xi, Pengcheng and Vald{\'e}s, Julio J and Wallace, Bruce and Goubran, Rafik},
  booktitle={2024 IEEE International Symposium on Medical Measurements and Applications (MeMeA)},
  pages={1--6},
  year={2024},
  organization={IEEE}
}

@inproceedings{selvaraju2017grad,
  title={Grad-cam: Visual explanations from deep networks via gradient-based localization},
  author={Selvaraju, Ramprasaath R and Cogswell, Michael and Das, Abhishek and Vedantam, Ramakrishna and Parikh, Devi and Batra, Dhruv},
  booktitle={Proceedings of the IEEE international conference on computer vision},
  pages={618--626},
  year={2017}
}

@article{Shen2024,
title = {Novel audio characteristic-dependent feature extraction and data augmentation methods for cough-based respiratory disease classification},
journal = {Comput. Biol. Med.},
volume = {179},
pages = {108843},
year = {2024},
author = {Jiakun Shen and Xueshuai Zhang and Yu Lu and Pengfei Ye and Pengyuan Zhang and Yonghong Yan}
}

@inproceedings{zeiler2014visualizing,
  title={Visualizing and understanding convolutional networks},
  author={Zeiler, Matthew D and Fergus, Rob},
  booktitle={European conference on computer vision},
  pages={818--833},
  year={2014},
  organization={Springer}
}

@inproceedings{ribeiro2016should,
  title={``{Why} should i trust you''' {Explaining} the predictions of any classifier},
  author={Ribeiro, Marco Tulio and Singh, Sameer and Guestrin, Carlos},
  booktitle={Proceedings of the 22nd ACM SIGKDD international conference on knowledge discovery and data mining},
  pages={1135--1144},
  year={2016}
}

@inproceedings{sanim2023identification,
  title={Identification of {COVID-19} from Other Upper Respiratory Tract Infections Using {Random Undersampling and LIME-based XAI Model}},
  author={Sanim, Mostofa Shariar and Hasan, Khan Mehedi and others},
  booktitle={2023 14th International Conference on Computing Communication and Networking Technologies (ICCCNT)},
  pages={1--6},
  year={2023},
  organization={IEEE}
}

@INPROCEEDINGS{wullenweber2022coughlime,
  author={Wullenweber, Anne and Akman, Alican and Schuller, Björn W.},
  booktitle={2022 44th Annual International Conference of the IEEE Engineering in Medicine and Biology Society (EMBC)},
  title={{CoughLIME: Sonified Explanations for the Predictions of {COVID}-19 Cough Classifiers}},
  year={2022},
  volume={},
  number={},
  pages={1342-1345},
  }

@ARTICLE{ramalingman2006,
  author={Ramalingam, Arunan and Krishnan, Sridhar},
  journal={IEEE Trans. Inf. Forensics Security},
  title={Gaussian Mixture Modeling of Short-Time Fourier Transform Features for Audio Fingerprinting},
  year={2006},
  volume={1},
  number={4},
  pages={457-463},
  }

@book{giannakopoulos2014introduction,
  title={Introduction to audio analysis: a MATLAB{\textregistered} approach},
  author={Giannakopoulos, Theodoros and Pikrakis, Aggelos},
  year={2014},
  publisher={Academic Press}
}

@article{li2020eeg,
  title={EEG-based intention recognition with deep recurrent-convolution neural network: Performance and channel selection by Grad-CAM},
  author={Li, Yurong and Yang, Hao and Li, Jixiang and Chen, Dongyi and Du, Min},
  journal={Neurocomputing},
  volume={415},
  pages={225--233},
  year={2020},
  publisher={Elsevier}
}

@INPROCEEDINGS{amado2024,
  author={Amado-Caballero, P. and Garmendia-Leiza, J. R. and Aguilar-García, M. D. and Fernández-Martínez-De-Septiem, C. and San-José-Revuelta, L. M. and García-Ruano, A. and Alberola-López, C. and Casaseca-De-La-Higuera, P.},
  booktitle={2024 46th Annual International Conference of the IEEE Engineering in Medicine and Biology Society (EMBC)},
  title={Audio Cough Analysis by Parametric Modelling of Weighted Spectrograms to Interpret the Output of Convolutional Neural Networks},
  year={2024},
  volume={},
  number={},
  pages={1-4},
  }




\end{document}